\documentclass[9pt,twocolumn,twoside]{pnas-new}
\setboolean{displaywatermark}{false}
\templatetype{pnasresearcharticle} % Choose template 
\usepackage{amsmath}
\usepackage{amssymb}
\usepackage{amsthm}
\usepackage{amsfonts}
\usepackage{verbatim}
\usepackage{listings}
\usepackage{enumitem}
\usepackage{latexsym}
\usepackage{ dsfont }
\usepackage{bm}
\usepackage{url}
\usepackage[all]{xy}
\usepackage{graphicx}
\usepackage{color}
\usepackage{mathtools}
\usepackage{eufrak}
\usepackage[percent]{overpic}
\usepackage{epsfig,slashed}
\usepackage{epstopdf}
\usepackage{lipsum}
\usepackage{float}
\usepackage{mathtools}
\usepackage{natbib}
\usepackage{subfigure}
\usepackage[mathscr]{euscript}
\usepackage{textcomp}
\usepackage{graphicx}
\usepackage{wasysym}

\makeatletter
\newsavebox\myboxA
\newsavebox\myboxB
\newlength\mylenA

\newcommand*\xoverline[2][0.75]{%
    \sbox{\myboxA}{$\m@th#2$}%
    \setbox\myboxB\null% Phantom box
    \ht\myboxB=\ht\myboxA%
    \dp\myboxB=\dp\myboxA%
    \wd\myboxB=#1\wd\myboxA% Scale phantom
    \sbox\myboxB{$\m@th\overline{\copy\myboxB}$}%  Overlined phantom
    \setlength\mylenA{\the\wd\myboxA}%   calc width diff
    \addtolength\mylenA{-\the\wd\myboxB}%
    \ifdim\wd\myboxB<\wd\myboxA%
       \rlap{\hskip 0.5\mylenA\usebox\myboxB}{\usebox\myboxA}%
    \else
        \hskip -0.5\mylenA\rlap{\usebox\myboxA}{\hskip 0.5\mylenA\usebox\myboxB}%
    \fi}
\makeatother

\newcommand{\mc}{\mathcal}

\newcommand{\bra}[1]{\langle {#1} |}
\newcommand{\ket}[1]{| {#1} \rangle}
\newcommand{\vect}[1]{\boldsymbol{#1}}

\definecolor{myblue}{rgb}{0.1, 0.2, 0.7}
\usepackage{pdfpages}
\usepackage{pgffor}
\makeatletter
\AtBeginDocument{\let\LS@rot\@undefined}
\makeatother

\DeclareGraphicsExtensions{.png}

\title{Quantum phases of Rydberg atoms\\ on a kagome lattice}

\author[a,1]{Rhine Samajdar}
\author[a,b]{Wen Wei Ho} 
\author[c,d]{Hannes Pichler}
\author[a]{Mikhail D. Lukin}
\author[a]{Subir Sachdev}

\affil[a]{Department of Physics, Harvard University, Cambridge, MA 02138, USA}
\affil[b]{Department of Physics, Stanford University, Stanford, CA 94305, USA}
\affil[c]{Institute for Theoretical Physics, University of Innsbruck, Innsbruck A-6020, Austria}
\affil[d]{Institute for Quantum Optics and Quantum Information, Austrian Academy of Sciences, Innsbruck A-6020, Austria}

\leadauthor{Samajdar} 

\significancestatement{Programmable quantum simulators based on Rydberg atom arrays have recently emerged as versatile platforms for exploring exotic many-body phases and quantum dynamics of strongly correlated systems. In this work, we theoretically investigate the quantum phases that can be realized by arranging such Rydberg atoms on a kagome lattice. Along with an extensive analysis of the states which break lattice symmetries due to classical correlations, we identify an intriguing new regime that constitutes a promising candidate for hosting a phase with long-range quantum entanglement and topological order. Our results suggest a novel route to experimentally realizing and probing highly entangled quantum matter. 
}

\authorcontributions{R.S. and S.S. conceived the research, R.S. performed the DMRG computations, W.W.H. and H.P. undertook the exact diagonalization studies, and H.P. and M.D.L. explored experimental implications and feasibility. All authors discussed the results and contributed to the manuscript.}
\authordeclaration{The authors declare no competing interests.}

\correspondingauthor{\textsuperscript{1}To whom correspondence should be addressed. E-mail: rhine\_samajdar@g.harvard.edu}


\begin{abstract}
We analyze the zero-temperature phases of an array of neutral atoms on the kagome lattice,  interacting via laser excitation to atomic Rydberg states. Density-matrix renormalization group calculations reveal the presence of a wide variety of complex solid phases with broken lattice symmetries. In addition, we identify a novel regime with
dense Rydberg excitations that has a large entanglement entropy and no local order parameter associated with lattice symmetries. From a mapping to the triangular lattice quantum dimer model, and theories of quantum phase transitions out of the proximate solid phases, we argue that this regime could contain one or more phases with topological order. Our results provide the foundation for theoretical and experimental explorations of crystalline and liquid states using programmable quantum simulators based on Rydberg atom arrays. 
\end{abstract}

\dates{Submitted to PNAS on July 28, 2020}

\begin{document}

\maketitle
\thispagestyle{firststyle}
\ifthenelse{\boolean{shortarticle}}{\ifthenelse{\boolean{singlecolumn}}{\abscontentformatted}{\abscontent}}{}

\dropcap{T}he search for quantum phases with fractionalization, emergent gauge fields, and anyonic excitations has been a central focus of research in quantum matter for the past three decades \cite{WenReview2017,SachdevReview2018}.  Such systems feature long-range many-body quantum entanglement, which can, in principle, be exploited for fault-tolerant quantum computing \cite{kitaev2003fault}. The best-studied examples in this regard are the fractional quantum Hall states found in high magnetic fields \cite{stormer1999fractional}. While such states have, by now, been realized in a wide variety of experimental systems, their intrinsic topological properties, including anyonic statistics, are challenging to detect and control directly \cite{Manfra2020}. 
In the absence of a magnetic field, the simplest anyonic phase compatible with time-reversal symmetry is the so-called $\mathbb{Z}_2$ spin liquid \cite{read1991large,wen1991}, which has the same topological order as the ``toric code'' \cite{kitaev2003fault}. While there are some indications that such a phase may be present in electronic systems on the kagome lattice \cite{YoungLee12,Imai15,ChineseLett17}, thus far, these quantum spin liquid (QSL) states have evaded direct experimental detection.  

In the search for QSLs, systems with frustration \cite{balents2010spin, savary2016quantum}---which can be either of geometric origin or induced by further-neighbor couplings---constitute a promising avenue of exploration. Motivated by this consideration,  here, we investigate many-body states of neutral atom arrays, interacting via laser excitation to atomic Rydberg states \cite{weimer2010rydberg}, that have been found to display a variety of interesting correlated quantum phases in one and two dimensions \cite{labuhn2016tunable,bernien2017probing,samajdar2018numerical,whitsitt2018quantum,keesling2019quantum,PhysRevLett.124.103601,de2019observation}. Specifically, we examine a realistic model of Rydberg atoms on the kagome lattice, and perform density-matrix renormalization group (DMRG) computations to establish its rich phase diagram as a function of laser parameters and atomic distances. These calculations reveal the formation of several intricate solid phases with long-range density-wave order. We show that one of these ordered phases actually emerges from a highly degenerate manifold of classical states via a quantum order-by-disorder mechanism. 
We also find a strongly correlated ``liquid regime'' of parameter space (identified by the star in Fig.~\ref{fig:PD}) where the density of Rydberg excitations is limited by the interactions, in contrast to the gas-like ``disordered regime'' where the laser driving induces independent atomic excitations. While for most interaction strengths, solid phases appear in such a dense regime, we observe that the liquid regime has no local order, and significant entanglement entropy. 
We employ a mapping to the triangular lattice quantum dimer model \cite{roychowdhury2015z}, which correctly describes the solid phases proximate to the liquid regime in the Rydberg model.  Theories for quantum phase transitions out of these solid phases then suggest that part of this liquid regime can host states with  long-range topological order. While our numerical results do not provide direct evidence for topological order over the system sizes studied,  we demonstrate that this regime should be readily accessible in experiments, raising the possibility of experimental investigations of entangled quantum matter. 
Remarkably, this is made possible simply using appropriate lattice geometries and innate interactions, even \textit{without} carefully engineering specific gauge constraints \cite{ZollerReview}.

\section*{Kagome lattice Rydberg model}

\begin{figure*}[b]
\includegraphics[width=\linewidth, trim={0.05cm 0 0.0cm 0}, clip]{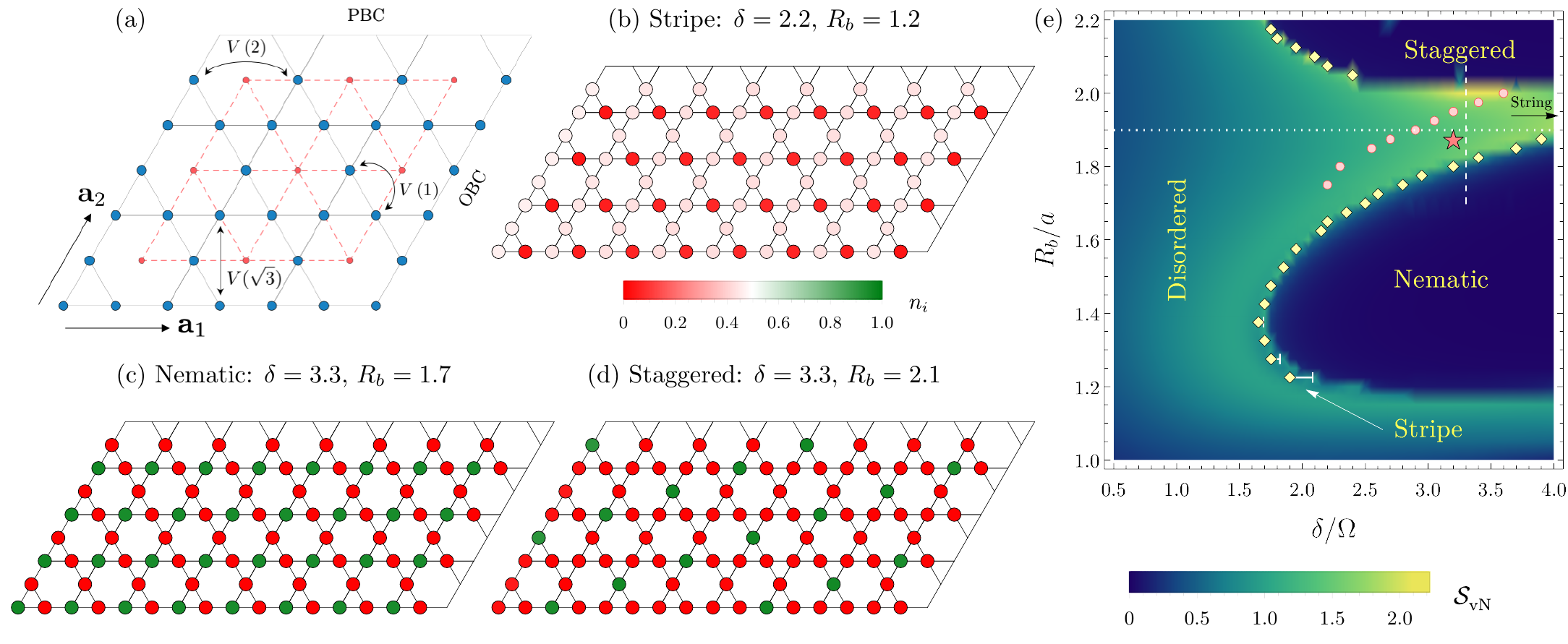}
\caption{\label{fig:PD}\textbf{Phases of the kagome lattice Rydberg atom array}. (a) Geometry of the kagome lattice; the lattice vectors are $\mathbf{a}_1$\,$=$\,$(2, 0)$, $\mathbf{a}_2$\,$=$\,$(1, \sqrt{3})$. Periodic (open) boundary conditions, designated by PBC (OBC), are imposed along the $\mathbf{a}_2$\,($\mathbf{a}_1$) direction, resulting in a cylinder. The blue dots are the sites of the original kagome lattice, where the atoms reside, while the red points outline the medial triangular lattice formed by connecting the centers of the kagome hexagons. (b--d) The various possible symmetry-broken ordered phases. Each lattice site is color coded such that green (red) signifies the atom on that site being in the Rydberg (ground) state. (e) Phase diagram of the Hamiltonian~(\ref{eq:Rydberg}) in the $\delta$-$R_b$ plane. The yellow diamonds and the pink circles are determined from the maxima of the susceptibility at each $R_b$; the former correspond to the finite-size pseudocritical points delineating the boundaries of the ordered phases. The white bars delimit the extent of the stripe phase. The string phase (see Fig.~\ref{fig:2/9}) lies at larger detuning, beyond the extent of this phase diagram, as conveyed by the black arrow. The correlated liquid regime is marked by a red star. The cuts along the dotted and dashed lines are analyzed in Figs.~\ref{fig:SVN} and \ref{fig:DCut}, respectively. }
\end{figure*}

Our interest lies in studying the phases of neutral atoms arranged on a kagome lattice, as sketched in Fig.~\ref{fig:PD}(a). Each kagome unit cell comprises three sites on a triangular scaffolding and the primitive vectors of this lattice are $\mathbf{a}_1$\,$=$\,$(2 a, 0)$ and $\mathbf{a}_2$\,$=$\,$(a, \sqrt{3} a)$, where the lattice constant $a$ is the spacing between two nearest-neighboring sites. Let us denote the number of complete unit cells along $\mathbf{a}_\mu$ by $N_\mu$. In a minimal model, each atom can be regarded as a two-level system with $\ket{g}_i$ and $\ket{r}_i$ representing the internal ground state and a highly excited Rydberg state of the $i$-th atom. The system is driven by a coherent laser field, characterized by a Rabi frequency, $\Omega$, and a detuning, $\delta$.
Putting these terms together, and taking into account the interactions between atoms in Rydberg states \cite{saffman2010quantum}, we arrive at the Hamiltonian 
\begin{alignat}{1}
\label{eq:Rydberg}
\nonumber H_{\rm Ryd}&=\sum_{i=1}^N \frac{\Omega}{2} \left(\ket{g}_i\!\bra{r}+\ket{r}_i\!\bra{g} \right)-\delta \ket{r}_i\!\bra{r} \\
&+ \frac{1}{2}\sum_{(i,j)} V\left(\lvert \lvert \vect{x}^{}_i-\vect{x}^{}_j \rvert  \rvert/a\right)\ket{r}_i\!\bra{r}\otimes \ket{r}_j\!\bra{r},
\end{alignat}
where the integers $i,j$ label sites (at positions $\vect{x}_{i,j}$) of the lattice, and the repulsive interaction potential is of the van der Waals form $V(R)$\,$=$\,$\mathcal{C}/R^6$ \cite{browaeys2016experimental}. Crucially, the presence of these interactions modifies the excitation dynamics. A central role in the physics of this setup is played by the phenomenon of the Rydberg blockade \cite{jaksch2000fast, lukin2001} in which  
strong nearest-neighbor interactions ($V(1)$\,$\gg$\,$|\Omega|$,\,$|\delta|$) can effectively prevent two neighboring atoms from simultaneously being in Rydberg states. The excitation of one atom thus inhibits that of another and the associated sites are said to be blockaded. By reducing the lattice spacing $a$, sites spaced further apart can be blockaded as well and it is therefore convenient to parametrize $H_{\rm Ryd}$ by the ``blockade radius'', defined by the condition $V (R_b/a) \equiv \Omega$ or equivalently, $\mc{C}$\,$\equiv$\,$\Omega\,R_b^6$.
Finally, we recognize that by identifying $|g \rangle$, $|r \rangle$ with the two states of a $S$\,$=$\,$1/2$ spin, 
$H_{\rm Ryd}$ can also be written as a quantum Ising spin model with $\mathcal{C}/R^6$ interactions in the presence of longitudinal ($\delta$) and transverse ($\Omega$) fields \cite{sachdev2002mott}.

We determine the quantum ground states of $H_\text{Ryd}$ for different values of $\delta/\Omega$ and $R_b/a$ using DMRG \cite{white1992density, white1993density}, which has been extensively employed on the kagome lattice to identify both magnetically ordered and spin liquid ground states of the antiferromagnetic Heisenberg model \cite{jiang2008density, yan2011spin, depenbrock2012nature}. The technical aspects of our numerics are documented in Sec.~I of the Supporting Information (SI). In particular, we work in the variational space spanned by matrix product state (MPS) ans\"{a}tze of bond dimensions up to $d$\,$=$\,$3200$. 
Although $(i, j)$ runs over all possible pair of sites in \eqref{eq:Rydberg}, this range is truncated in our computations, where we retain interactions between atoms separated by up to $2a$ (third-nearest neighbors), as shown in Fig.~\ref{fig:PD}(a). In order to mitigate the effects of the boundaries, we place the system on a cylindrical geometry by imposing open (periodic) boundary conditions along the longer (shorter) $\mathbf{a}_1$\,($\mathbf{a}_2$)-direction. The resulting cylinders are labeled by the direction of periodicity and the number of sites along the circumference; for instance, Fig.~\ref{fig:PD}(a) depicts a YC6 cylinder. Since the computational cost of the algorithm (for a constant accuracy) scales exponentially with the width of the cylinder \cite{liang1994approximate}, here, we limit the systems considered to a maximum circumference of 12 lattice spacings. Unless specified otherwise, we always choose the linear dimensions $N_1, N_2$ so as to yield an aspect ratio of $N_1/ N_2 \simeq 2$, which is known to minimize finite-size corrections and optimize DMRG results in two dimensions \cite{white2007neel, stoudenmire2012studying}. 

\section*{Phase diagram}

We first list the various phases of the Rydberg Hamiltonian that can arise on the kagome lattice. Without loss of generality,  we set $\Omega = a=1$ hereafter for notational convenience. At large negative detuning, it is energetically favorable for the system to have all atoms in the state $\rvert g \rangle$, corresponding to a trivial ``disordered'' phase with no broken symmetries \cite{nikolic2005theory}. As $\delta/\Omega$ is tuned towards large positive values, the fraction of atoms in $\rvert r \rangle$ increases but the geometric arrangement of the excitations is subject to the constraints stemming from the interactions between nearby Rydberg atoms.
This competition between the detuning and the previously identified blockade mechanism results in so-called ``Rydberg crystals'' \cite{pohl2010dynamical}, in which Rydberg excitations are arranged regularly across the array, engendering symmetry-broken density-wave ordered phases \cite{PhysRevLett.124.103601}. On the kagome lattice, the simplest such crystal that can be formed---while respecting the blockade restrictions---is constructed by having an atom in the excited state on exactly \textit{one} out the three sublattices in the kagome unit cell. This is the ordering pattern of the ``nematic'' phase [Fig.~\ref{fig:PD}(c)], which is found in a regime where only nearest-neighbor sites are blockaded. The nematic order spontaneously breaks the threefold rotational ($C_3$) symmetry of the underlying kagome lattice, so, for an infinite system, the true ground state is triply degenerate within this phase. Even though $H_{\rm Ryd}$ does not conserve the number of Rydberg excitations, the ordered state can still be characterized by a ``filling fraction'' upon taking the classical limit $\delta/\Omega \rightarrow \infty$, $R_b/a \ne 0$, which, in this case, leads to a density of $\langle n_i \rangle$\,$=$\,$1/3$, where $n_i \equiv \ket{r}_i\bra{r}$. 

Curiously, the nematic phase is separated from the trivial disordered one by a sliver of a quantum solid without any classical analogue, to wit, the stripe phase seen in Fig.~\ref{fig:PD}(b). This state also breaks the $C_3$ symmetry; accordingly, between the disordered and stripe phases, one encounters a $\mathbb{Z}_3$-symmetry-breaking quantum phase transition (QPT) \cite{sachdev2011quantum} in the universality class of the (2+1)D three-state Potts model \cite{janke1997three}, while the QPT demarcating stripe and nematic is first-order. Although both phases break the same symmetry, the stripe ordering is distinguished from the nematic by a substantial and equal density on \textit{two} sublattices of the unit cell. The formation of these stripes can be attributed to quantum fluctuations \cite{PhysRevLett.124.103601}, which help stabilize the phase in a narrow window as follows. The system optimizes the geometric packing in a configuration where all atoms on one sublattice are in the ground state, whereas those on the other two sublattices are each in a quantum superposition formed by the ground state with a coherent admixture of the Rydberg state. These ``dressed'' atoms assist in offsetting the energetic penalty due to the interactions, while simultaneously maximizing the excitation density and therefore, the reduction in energy from $\delta$. The ensuant average density in the stripe phase is also $\langle n_i \rangle$\,$\sim$\,$1/3$, which explains its existence as a precursor to the nematic ordering. The extent of this phase narrows significantly with increasing $R_b$, so it is difficult to ascertain whether the transition between the lattice nematic and disordered phases is always a two-step one with the stripe order intervening. Nevertheless, based on our current data (see also Fig.~\ref{fig:DCut}), we believe it is likely that the stripe phase terminates at a tricritical point near the tip of the nematic dome instead of surrounding it throughout.
  
Proceeding to larger blockade radii, we find that the kagome Rydberg array hosts yet another solid phase with density-wave ordering, namely the ``staggered'' phase [Fig.~\ref{fig:PD}(d)]. This phase, which bears a twelvefold ground-state degeneracy, is realized when interactions between neighboring Rydberg atoms are sufficiently strong enough to blockade third-nearest-neighbor sites, so the excitations are positioned a distance of $\sqrt{7}$ apart.
The resultant Rydberg crystals are formed of a 12-site unit cell with lattice vectors $4\mathbf{a}_1$ and $2\mathbf{a}_1 + \mathbf{a}_2$; the associated classical density is $1/6$. The staggered phase remains stable up to $R_b$\,$\lesssim$\,$\sqrt{7}$, beyond which fourth-nearest neighbors are also blockaded.

\begin{figure}[tb]
\includegraphics[width=\linewidth]{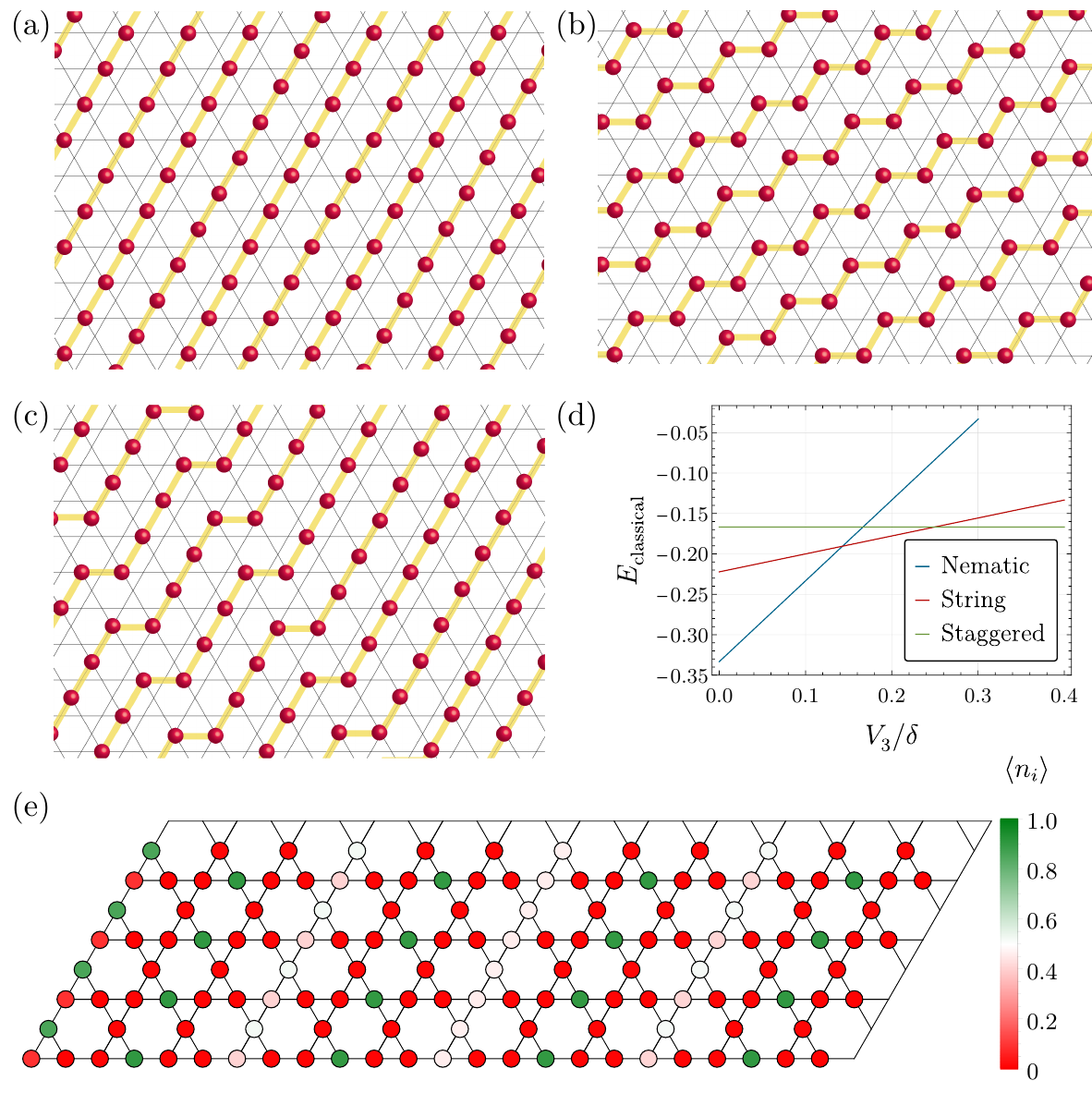}
\caption{\label{fig:2/9}\textbf{Crystalline phase at $2/9$ filling on the kagome lattice}. (a--c) Classically ordered states at $f$\,$=$\,$2/9$; while we sketch only three configurations here, the number of such states---with the same filling fraction---scales exponentially with the system size. The Rydberg excitations (red) are arranged in ``strings'' (yellow) that span the lattice. (d) Comparing the three possible classical phases, we find that the energy (at $\Omega$\,$=$\,$0$) is minimized by the string-ordered state over a finite region between the nematic and the staggered phases. (e) Rydberg crystal formed in the string phase at $\delta$\,$=$\,$4.00$, $R_b$\,$=$\,$1.95$ on a wide ($N_1$\,$=$\,$12$) YC8 cylinder.}
\end{figure}

Equipped with the information above, we now turn to assembling the full phase diagram of $H_{\rm{Ryd}}$.
An unbiased diagnostic to do so is the bipartite von Neumann entanglement entropy (EE) of the ground state $\mc{S}_{\rm{vN}}$\,$\equiv$\,$-\mathrm{Tr}\,(\rho_r \ln \rho_r)$, $\rho_r$ being the reduced density matrix for each subsystem when the cylinder is partitioned in half along $\mathbf{a}_1$. On going from the disordered phase to an ordered one, $\mc{S}_{\rm{vN}}$ gradually increases, peaks near the quantum critical point (QCP), and then drops sharply inside the solid phase [see also Fig.~\ref{fig:SVN}(c)]. This is because DMRG prefers states with low entanglement and systematically converges to a so-called Minimal Entropy State (MES) \cite{jiang2012identifying, stoudenmire2012studying}, which is simply one of the symmetry-broken states rather than their superposition. This drastic decline in $\mc{S}_{\rm{vN}}$ traces out the two lobes seen in Fig.~\ref{fig:PD}(e), which mark the phase boundaries of the nematic and staggered orders. In the limit of large detuning, there is another density-wave ordered phase between these two lobes, which we christen the ``string'' phase and discuss next.

\section*{Quantum order-by-disorder}

In the classical limit of $\delta/\Omega \rightarrow \infty$, the periodic arrangement of Rydberg excitations (or equivalently, hard-core bosons) on the kagome lattice can result in additional ordered phases besides the nematic and the staggered at various fractional densities \cite{huerga2016staircase}. To see this, one can simply minimize the \textit{classical} energy, which is determined solely by the competition between the detuning and the repulsive interactions.
In the parameter range of interest ($R_b$\,$\lesssim$\,$2.25$), it is not difficult to observe [Fig.~\ref{fig:2/9}(d)] that this optimization yields three regions characterized by classical filling fractions of 
\begin{equation}
f = 
\begin{cases}
1/3;\quad V_3/\delta < 1/7,\\
2/9;\quad 1/7< V_3/\delta < 1/4,\\
1/6;\quad 1/4< V_3/\delta,
\end{cases}
\end{equation}
where $V_3$ represents the strength of the third-nearest-neighbor interactions. Since we have (temporarily) set $\Omega$\,$=$\,$0$, the ratio $V_3/\delta$ is the only independent tuning parameter for the Hamiltonian in this limit.

The phases at fillings of a third and a sixth can be readily identified as (the classical versions of) the familiar nematic and staggered orders [Figs.~\ref{fig:PD}(c) and (d)], respectively. In between the two, the system favors a separate highly degenerate classical ground state, forming what we dub the ``string'' phase. A few of the possible ordering patterns for a crystal belonging to this phase, with a filling fraction of $f$\,$=$\,$2/9$, are presented in Figs.~\ref{fig:2/9}(a--c). The arrangement of the Rydberg excitations resembles strings---which may be straight or bent---that stretch across the lattice. Interestingly, there are a macroscopic number of such states, all with the same classical energy, and this degeneracy grows exponentially with the linear dimensions of the system. For example, in Fig.~\ref{fig:2/9}(a), the positions of all the atoms in the Rydberg state can be uniformly shifted by $\pm\mathbf{a}_2/2$ for every other string without affecting the energy, leading to $\mc{O}(2^{N_1})$ potential configurations. Similarly, when the strings are bent, like in Fig.~\ref{fig:2/9}(c), there are $\mc{O}(N_2)$ locations where a kink can be formed, and correspondingly, $\mc{O}(2^{N_2})$ states of this type.

The large classical degeneracy raises the question of the fate of this phase once we reinstate a nonzero transverse field, $\Omega$. There are two natural outcomes to consider. Firstly, a superposition of the classical ground states can form a quantum liquid with topological order, as is commonly seen to occur in quantum dimer models \cite{Moessner2011}. However, a necessary condition in this regard is the existence of a \textit{local} operator which can connect one classical ground state with another. Since the individual ground states are made up of parallel strings, they are macroscopically far away from each other, and it would take an operator with support of the size of the system length to move between different classical configurations, thus violating the requirement of locality. This brings us to the second possibility, namely, that a quantum ``order-by-disorder'' phenomenon \cite{villain1980order,shender1982antiferromagnetic} prevails. In this mechanism,  quantum fluctuations lower the energy of particular classical states from amongst the degenerate manifold; the system then orders in a state around which the cost of excitations is especially cheap. In this case, one could anticipate a string-ordered solid phase, which should be easily identifiable from the structure factor.

The DMRG numerics confirm our intuition that such a crystal should emerge in the phase diagram at sufficiently high detunings. On the YC8 cylinder with $N_1$\,$=$\,$8$, the string phase appears at detunings beyond the range rendered in Fig.~\ref{fig:PD}(e). However, it is manifestly observed, for a wider geometry, in Fig.~\ref{fig:2/9}(e), which illustrates the local magnetizations inside the string phase (at $\delta$\,$=$\,$4.00$, $R_b$\,$=$\,$1.95$) on a YC8 cylinder of length $N_1$\,$=$\,$12$ (chosen so as to be fully compatible with the string order). The ground state found by finite DMRG is patently ordered with the system favoring a configuration of straight strings that wrap around the cylinder, thereby lifting the macroscopic classical degeneracy. This is in contrast to the expectation from naive second-order perturbation theory, which picks out the maximally kinked classical state.

\section*{Signatures of density-wave orders}

In totality, we have thus detected four solid phases on the kagome lattice. All these ordered states  can be identified from either their respective structure factors, or the relevant order parameters, as we now show.

\begin{figure}[tb]
\includegraphics[width=\linewidth]{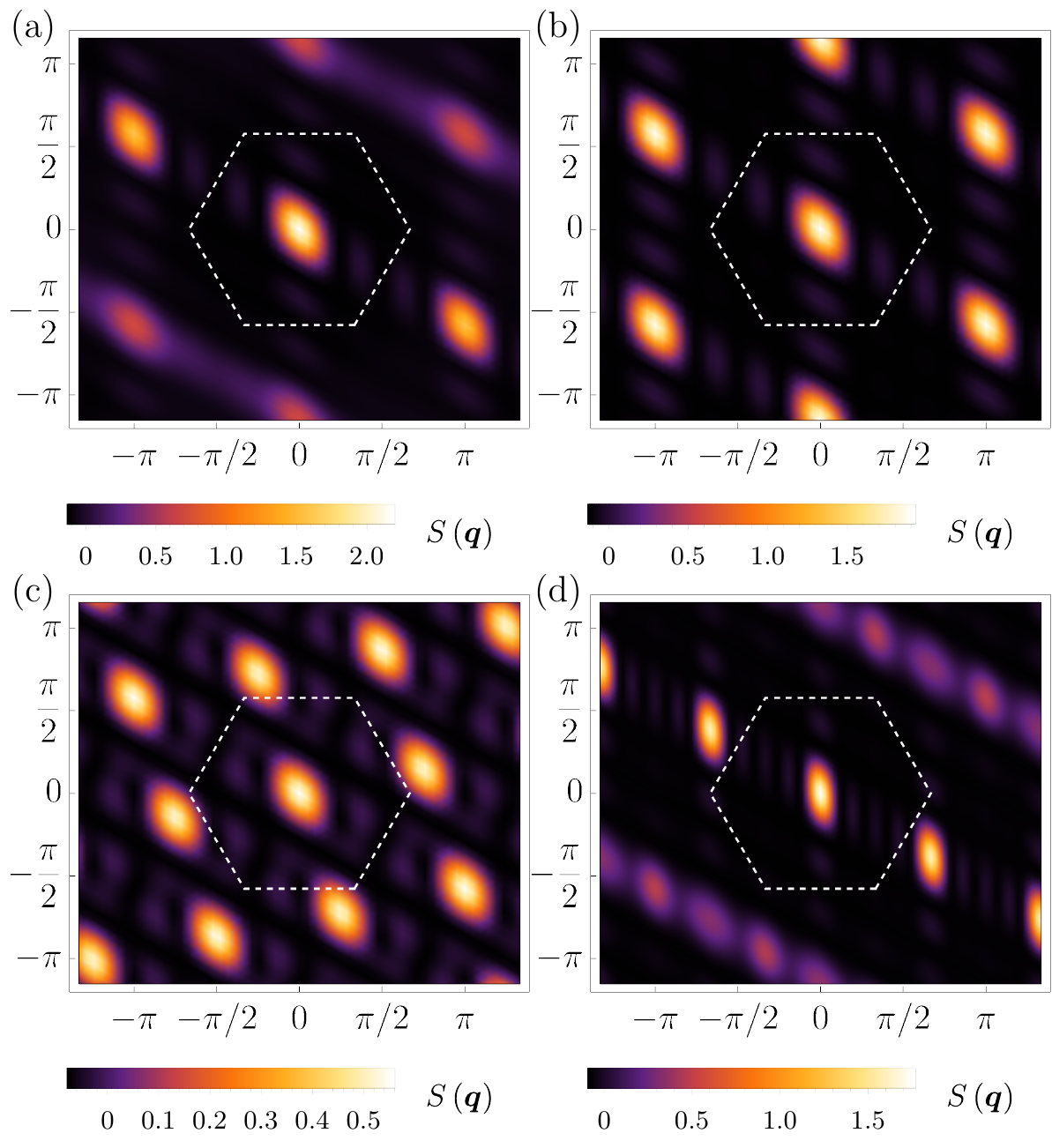}
\caption{\label{fig:SF}\textbf{Static structure factors of the various ordered phases.} $S(\vect{q})$ displays pronounced and well-defined peaks for the (a) stripe ($\delta$\,$=$\,$2.20$, $R_b$\,$=$\,$1.20$), (b) nematic ($\delta$\,$=$\,$3.30$, $R_b$\,$=$\,$1.70$), (c) staggered ($\delta$\,$=$\,$3.30$, $R_b$\,$=$\,$2.10$), and (d) string ($\delta$\,$=$\,$4.00$, $R_b$\,$=$\,$1.95$) orders. The dashed white hexagon marks the first Brillouin zone of the kagome lattice. The structure factor for the string phase is computed on the cylindrical geometry shown in Fig.~\ref{fig:2/9}(e).}
\end{figure}

With a view to extracting bulk properties, in the following, we work with the central half of the system that has an effective size of $N_c$\,$=$\,$3 N_2^2$. Evidence for ordering or the lack thereof can be gleaned from the static structure factor, which is the Fourier transform of the instantaneous real-space correlation function
\begin{equation}
\label{eq:SF}
S (\vect{q}) = \frac{1}{N_c}\sum_{i,j} e^{{\rm i} \vect{q}\cdot(\vect{x}_i-\vect{x}_j)} \langle n^{}_i n^{}_j  \rangle
\end{equation}
with the site indices $i$,\,$j$ restricted to the central $N_2$\,$\times$\,$N_2$ region of the cylinder. At a blockade radius of $R_b$\,$=$\,$1.7$, which stations one in the nematic phase [Fig.~\ref{fig:SF}(b)], the structure factor has pronounced maxima at the corners of the (hexagonal) extended Brillouin zone, occurring at $\vect{Q}$\,$=$\,$\pm \mathbf{b}_1$, $\pm \mathbf{b}_2$, $\pm (\mathbf{b}_1$\,$+$\,$\mathbf{b}_2)$, where $\mathbf{b}_1$\,$=$\,$(\pi, -\pi/\sqrt{3})$ and $\mathbf{b}_2$\,$=$\,$(0, 2\pi/\sqrt{3})$ are the reciprocal lattice vectors. A subset of these maxima also persists for the stripe phase [Fig.~\ref{fig:SF}(a)]---this is in distinction to the nematic phase wherein the peaks at all six reciprocal lattice vectors are of equal strength. In the presence of staggered ordering [Fig.~\ref{fig:SF}(c)], the peaks are comparatively weaker but prominent nonetheless, appearing at $\vect{Q}$\,$=$\,$\pm \mathbf{b}_1$, $\pm(\pi/2, \sqrt{3}\pi/2)$, $\pm(3\pi/4, \pi/(4 \sqrt{3}))$, and $\pm(-\pi/4, 5\pi/(4 \sqrt{3}))$. Likewise, in the string phase [Fig.~\ref{fig:SF}(d)], conspicuous maxima are seen to occur at $\pm 2\mathbf{b}_1/3$ for the ground-state configuration where straight strings encircle the lattice. While we list here the ordering wavevectors for a finite system, let us briefly note that on an infinite lattice, the structure factors, of course, would additionally include $C_3$-rotated copies of the above.

One can also directly look at the order parameters that diagnose the possible symmetry-broken ordered states. For the nematic phase, an appropriate definition is 
\begin{equation*}
\Phi= \frac{3}{N_c} \left( \sum_{i\in \mathrm{A}}n_i+ \omega \sum_{i\in \mathrm{B}}n_i+\omega^2\sum_{i\in \mathrm{C}}n_i\right),
\end{equation*} where $\omega \equiv \exp(2\pi i/3)$ is the cube root of unity, and $\mathrm{A}$,\,$\mathrm{B}$,\,$\mathrm{C}$ denote the three sublattices of the kagome lattice. Similarly, in the staggered and string phases, one can define the (squared) magnetic order parameter $M_{\vect{Q}}^2$\,$\equiv$\,$S(\vect{Q})$, with $\vect{Q}$ chosen from among the observed peaks of the structure factor. These order parameters are more quantitatively addressed in Fig.~\ref{fig:DCut}(a), which catalogs the ground-state properties calculated at a fixed detuning of $\delta$\,$=$\,$3.3$ [dashed line in Fig.~\ref{fig:PD}(e)]; in particular, we observe that the nematic and staggered order parameters assume nontrivial values in exactly the regions predicted by the phase diagram.

\section*{Mapping to triangular lattice quantum dimer models}

At large detuning, we can approximately map the Rydberg system to a model of hard-core bosons at filling $f$ on the kagome lattice. The bosonic system \cite{BFG02,isakov2006spin,Melko2011,roychowdhury2015z,Meng18,Krempa20} has an extra $\mathrm{U}$(1) symmetry, which can be spontaneously broken in a superfluid phase; in the Rydberg model without the U(1) symmetry, the disordered phase is the counterpart of the superfluid. However, any nonsuperfluid topological states of the boson model are insensitive to the $\mathrm{U}$(1) symmetry, and can also be present in the Rydberg model. 

\begin{figure}[tb]
\includegraphics[width=\linewidth]{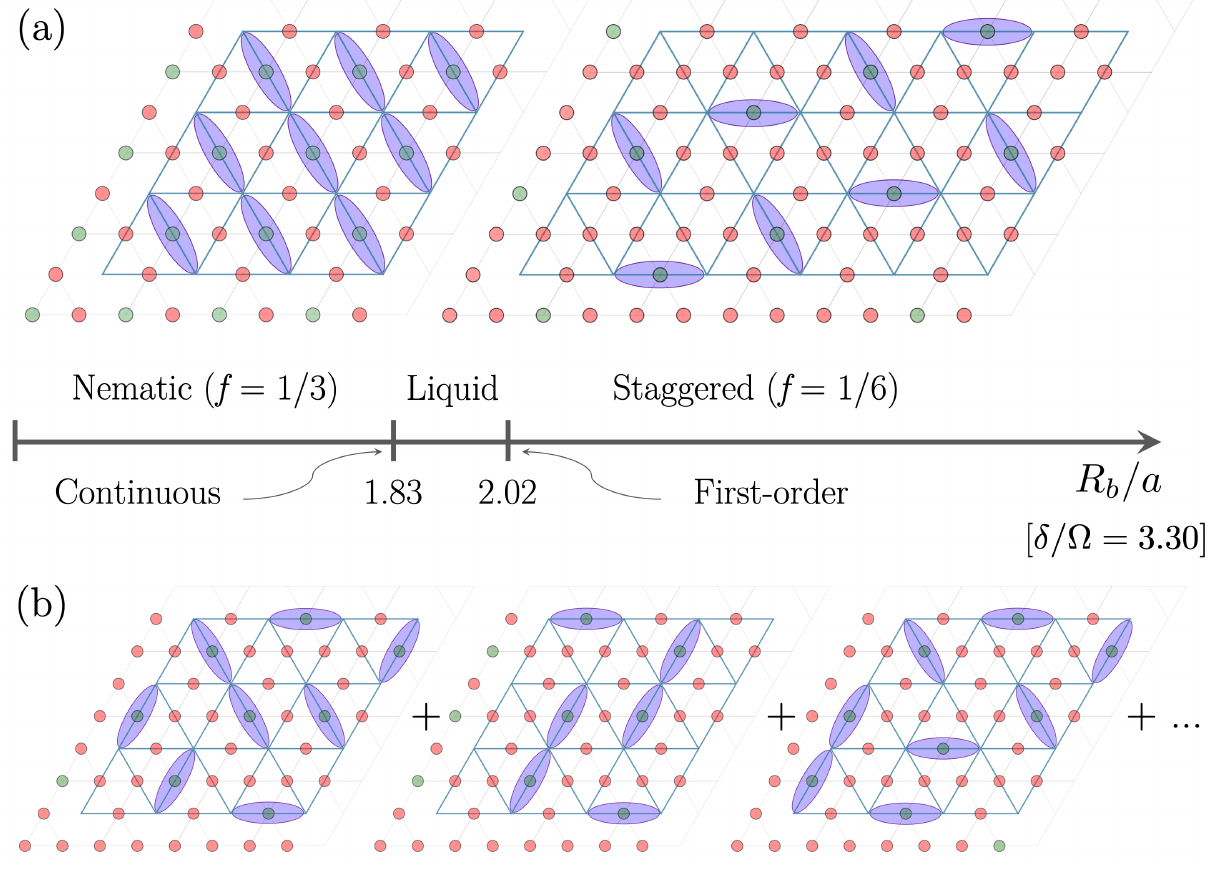}
\caption{\label{fig:dimer}\textbf{Correspondence between the Rydberg and quantum dimer models.} In the limit of large detuning, the Rydberg excitations can be mapped to a system of hard-core bosons, upon identifying each atom in state $\ket{r}$ ($\ket{g}$) as an occupied (empty) bosonic mode \cite{sachdev2002mott}. (a) The resultant boson model is at a filling of $f$\,$=$\,$1/3$ ($1/6$) for the nematic (staggered) phase. A boson on any \textit{site} of the kagome lattice (red/green points) can now be uniquely associated with a dimer on the corresponding \textit{bond} of the medial triangular lattice (blue lines) \cite{roychowdhury2015z}. The liquid regime is separated from the nematic (staggered) phase by a continuous (first-order) QPT. (b) Schematic depiction of a potential Rydberg liquid as a superposition of dimer configurations; note that, unlike in the QDM, the total number of dimers can fluctuate in the Rydberg model. }
\end{figure}

In the limit of strong interactions, hard-core bosons at filling $f$\,$=$\,$(1/2, 1/3, 1/6)$ on the \textit{kagome} lattice  map \cite{BFG02,isakov2006spin,Melko2011,roychowdhury2015z} onto an (odd, even, odd) quantum dimer model (QDM)  \cite{moessner2001resonating, moessner2001ising,ZiYang2020dimers} on the medial \textit{triangular} lattice with $N_d$\,$=$\,$(3, 2, 1)$ dimers per site, with odd/even referring to the parity of $N_d$. The triangular lattice of the QDM is formed by joining the centers of the kagome hexagons, and this correspondence is sketched in Fig.~\ref{fig:dimer}, which schematically shows the mapping between the different Rydberg solids and the phases of the QDM. A key observation here is that both solid phases next to the liquid regime (marked by the star in Fig.~\ref{fig:PD}) are {\it also\/} phases of the QDM: the nematic phase was found in the QDM with $N_d$\,$=$\,$2$ by Roychowdhury \textit{et al.} \cite{roychowdhury2015z}, and the staggered phase is present in the QDM with $N_d$\,$=$\,$1$ \cite{moessner2001resonating, moessner2001ising}. In both cases, a $\mathbb{Z}_2$ spin liquid phase with topological order has been found adjacent to these solid phases \cite{roychowdhury2015z,moessner2001resonating, moessner2001ising} in the QDMs.
Making the reasonable assumption that a QDM description for the Rydberg system holds in the vicinity of the phase boundaries of these solid states, we expect $\mathbb{Z}_2$ topological order in the liquid regime of the Rydberg model in Fig.~\ref{fig:PD}, proximate to the nematic and staggered solid phases. 

There is a subtle difference between the $\mathbb{Z}_2$ spin liquids found in the $N_d=1,2$ QDMs: the anyonic ``vison'' excitation picks up a Berry phase of $\pi$ ($2 \pi$) upon adiabatic transport around a site
of an odd (even) QDM.
\cite{jalabert1991spontaneous, sachdev1992kagome,sachdev1999translational,SachdevReview2018,senthil2000z,moessner2001resonating,moessner2001ising,MSF02,Essin:2013rca,QiFu15,ZV15}.
This distinction changes the projective symmetry group of the visons, and also holds for the $\mathbb{Z}_2$ spin liquids expected in the Rydberg model, which must therefore be odd/even as well. Consequently, the spin liquids proposed to be proximate to the staggered and nematic phases are not identical; one or both of them could be present in the liquid regime. 
Moreover, the vison Berry phase places important constraints on the non-topological states obtained by condensing visons: for instance, an odd $\mathbb{Z}_2$ spin liquid cannot have a vison-condensing phase transition to a trivial ``disordered'' state with no broken lattice symmetry, which is a manifestation of the Lieb-Schultz-Mattis theorem.

Roychowdhury \textit{et al.} \cite{roychowdhury2015z} studied the transition from the even $\mathbb{Z}_2$ spin liquid into the nematic phase. 
The visons in this $\mathbb{Z}_2$ spin liquid have an energy dispersion with minima at $\vect{M}_1$\,$=$\,$(\pi/2, -\pi/(2\sqrt{3}))$\,$=$\,$\mathbf{b}_1/2$, $\vect{M}_2$\,$=$\,$(0, \pi/\sqrt{3})$\,$=$\,$\mathbf{b}_2/2$ \cite{roychowdhury2015z}, and their condensation leads to the nematic ordering for which the dominant wavevectors are $\mathbf{b}_1$, $\mathbf{b}_2$ [Fig.~\ref{fig:SF}(b)]. The critical theory for this transition is an O$(3)$ Wilson-Fisher theory with cubic anisotropy \cite{roychowdhury2015z}, and this conclusion holds both for the QDM and the Rydberg system. It is interesting to compare this result to that for the transition from the nematic phase to the disordered phase of the Rydberg model, which was mentioned above to be in the universality class of the (2+1)D three-state Potts model and hence, weakly first-order \cite{janke1997three}. 
Therefore, the nematic phase can melt either by a first-order transition to a trivial disordered phase, or via a second-order one into a topological phase by fractionalizing the nematic order parameter.
So, the observation of a continuous O(3) transition out of the nematic phase to a phase without symmetry breaking would constitute nontrivial evidence for the presence of $\mathbb{Z}_2$ topological order in the latter. An apparent second-order transition in the nematic order parameter can be seen in Fig.~\ref{fig:DCut}(b) below, although our numerical accuracy is not sufficient to determine its universality class.

The transition from the staggered phase to the odd $\mathbb{Z}_2$ spin liquid of the $N_d$\,$=$\,$1$ QDM is first-order \cite{moessner2001resonating, moessner2001ising}, and we expect it to be so for the Rydberg model too.
This is compatible with the rapid increase of the staggered order parameter out of the liquid regime shown in Fig.~\ref{fig:DCut}(b). We also note that
the density of Rydberg excitations in the liquid regime ($\sim 0.2$) is close to that of the odd QDM ($f$\,$=$\,$1/6$). 

For both the even and odd $\mathbb{Z}_2$ spin liquids proposed for the liquid regime of the Rydberg model, there should be a sharp transition to the disordered phase described by the condensation of the bosonic $e$ anyons. Such a transition is not present in the QDMs, because the $e$ excitations have been projected out by the dimer constraint. This QPT is in the universality class of the Ising$^{*}$ Wilson-Fisher conformal field theory \cite{CSS94,schuler2016universal,whitsitt2016transition}, and can, in principle, be accessible in our system. However, we do not find clear-cut numerical evidence for it below, for our range of system sizes.

Extending the mapping from the Rydberg model to the QDM further, in Sec.~III of the SI, we compute the parameters in $(\delta, R_b)$-space where a QSL phase might be expected to exist for the Rydberg system based on the (previously known) regime of stability of the QDM spin liquid \cite{moessner2001resonating, moessner2001ising}. This calculation leads to an estimate of ($\delta$\,$=$\,$2.981$, $R_b$\,$=$\,$1.997$), which places us within the liquid regime of our phase diagram.

\section*{The liquid regime}

At moderately large values of the detuning, we find an intermediate correlated regime---designated by the red star in Fig.~\ref{fig:PD}(e)---which lies in between two solid phases but resists categorization as either. The nomenclature ``liquid'', as defined earlier, connotes that the Rydberg excitations form a dense state in which the blockade introduces significantly more correlations than in the disordered regime. Prompted by the considerations described in  the previous section, 
we first attempt  to uncover the existence of any phase transitions in the vicinity of this regime. To that end, we temporarily focus on a specific blockade radius, $R_b$\,$=$\,$1.9$ [dotted white line in Fig.~\ref{fig:PD}(e)], and look at variations of the ground-state properties along this one-dimensional cut. 

\begin{figure}[tb]
\includegraphics[width=\linewidth, trim={0.1cm 0.1cm 0 0}, clip]{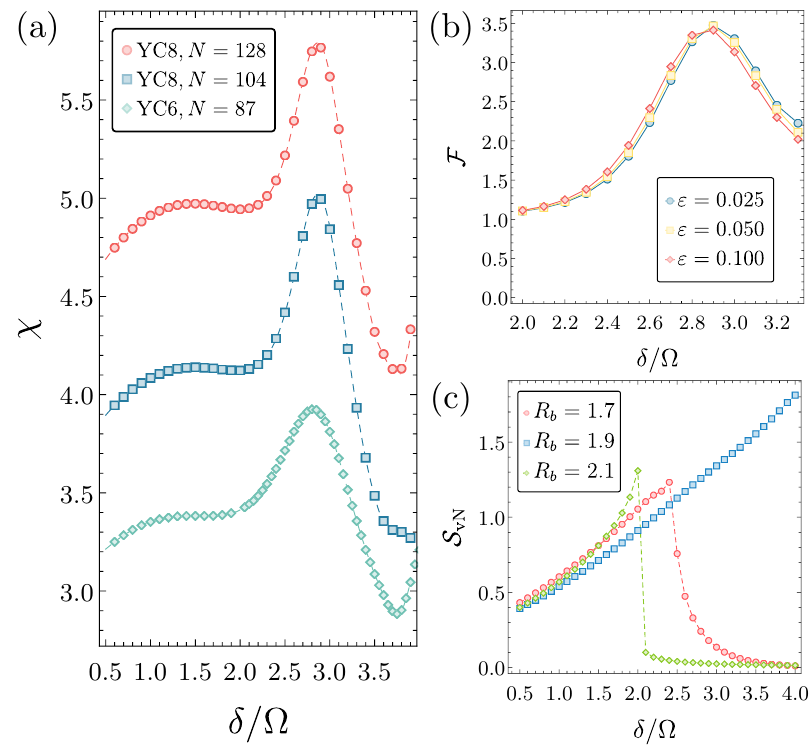}
\caption{\label{fig:SVN}\textbf{Signatures of a crossover into the liquid regime.} Along a line where the blockade radius is held constant at $R_b$\,$=$\,$1.90$, both (a) the susceptibility $\chi$ and (b) the fidelity susceptibility $\mc{F}$ exhibit a single peak at $\delta \approx 2.90$. (c) The behavior of the EE over the same detuning range, however, is distinct from the sharp drop observed across the QPTs into any of the ordered phases. On going to higher $\delta$, the system eventually transitions into either the nematic or the string phase depending on the blockade radius (or potentially, the boundary conditions). 
}
\end{figure}

The first such observable is the susceptibility, defined as the second derivative of the ground-state energy, $E_0$, with respect to the detuning, i.e., $\chi$\,$=$\,$- \partial^2 E_0 /\partial \,\delta^2$. On finite systems, the maxima of the susceptibility can often be used to identify possible QCPs, which are slightly shifted from their locations in the thermodynamic limit. In particular, for $R_b$\,$=$\,$1.9$, $\chi$ is plotted in Fig.~\ref{fig:SVN}(a), where a single peak in the response is visible at approximately $\delta$\,$=$\,$2.9$. This susceptibility peak---which is recorded by the pink circles in Fig~\ref{fig:PD}(e)---is also reproduced in exact diagonalization calculations on a 48-site torus (refer to Sec.~IV of the SI).

\begin{figure}[tb]
\includegraphics[width=0.99\linewidth, trim =0 0.1cm 0 0, clip]{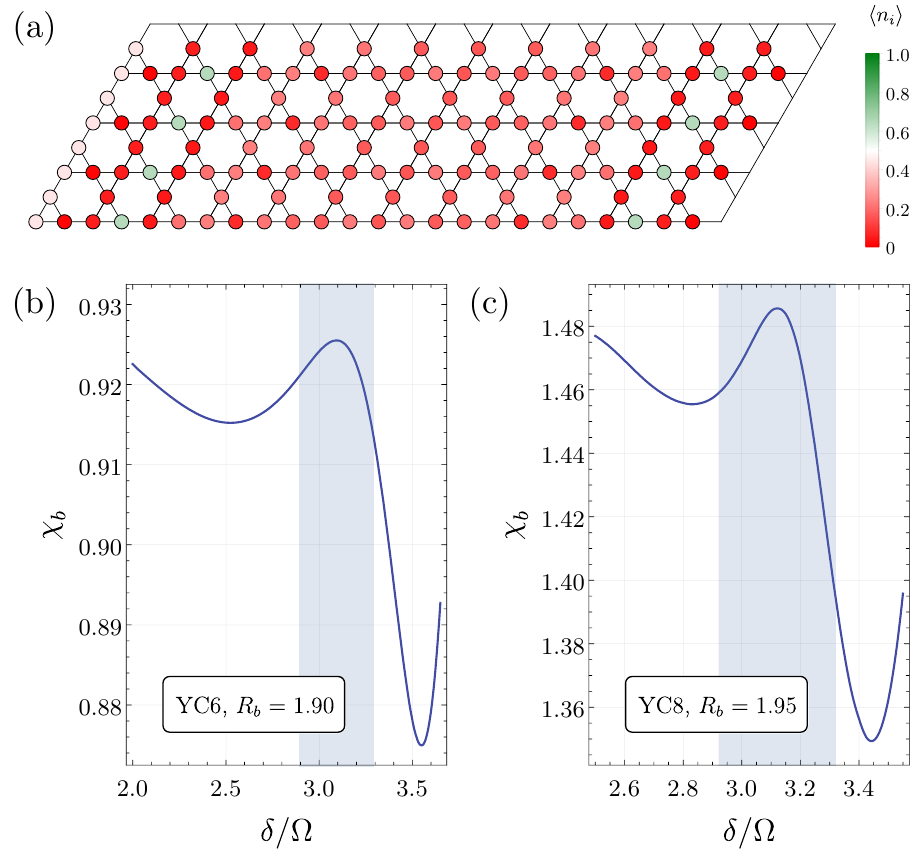}
\caption{\label{fig:Edge}\textbf{Disentangling bulk and boundary behaviors.} (a) Within the liquid regime---depicted here at $\delta$\,$=$\,$3.50$, $R_b$\,$=$\,$1.95$---the real-space magnetization profile communicates the absence of density-wave order; note that the edge-induced ordering does \textit{not} permeate into the bulk, which remains uniform. (b,c) Bulk susceptibilities: by construction, $\chi^{}_{b}$ should be insensitive to edge effects. In the left panel (b), $\chi^{}_{b}$ is determined from the second derivative of the difference between the energies of two YC6 cylinders with lengths $N_1$\,$=$\,$12$ and $N_1$\,$=$\,$9$. As in Fig.~\ref{fig:SVN}(a), with $R_b$ set to $1.90$, a clear local maximum appears at $\delta$\,$\approx$\,$3.09$, heralding the liquid regime. On YC8 cylinders (c), the bulk susceptibility, shown here along $R_b$\,$=$\,$1.95$, is calculated by applying the subtraction method to two systems of lengths $N_1$\,$=$\,$12$ and $N_1$\,$=$\,$8$. }
\end{figure}

A similar signature can be discerned in the quantum fidelity $\lvert \langle \Psi_0 (\delta) \vert  \Psi_0 (\delta+\varepsilon) \rangle \rvert$ \cite{cozzini2007quantum,zhou2008fidelity}, which measures the overlap between two ground-state wavefunctions $\Psi_0$ computed at parameters differing by $\varepsilon$. The fidelity serves as a useful tool in studying QPTs because, intuitively, it quantifies the similarity between two states, while QPTs are necessarily accompanied by an abrupt change in the structure of the ground-state wavefunction \cite{gu2010fidelity}. Zooming in on a narrower window around the susceptibility peak, we evaluate the fidelity susceptibility \cite{you2007fidelity}, which, in its differential form, is given by
\begin{equation}
\mc{F}  \equiv 2 \left[ \frac{1 - \lvert \langle \Psi_0 (\delta) \vert  \Psi_0 (\delta+\varepsilon) \rangle \rvert}{\varepsilon^2} \right].
\end{equation}
The fidelity susceptibility also displays a local maximum at $\delta$\,$\approx$\,$2.9$, indicating some change in the nature of the ground state as we pass into the liquid regime.
Unlike the QPTs into the ordered phases, the EE [Fig.~\ref{fig:SVN}(c)] does not drop as we cross this point but rather, continues to increase; however, its first derivative is nonmontonic at $\delta \approx 2.9$. This suggests that the final liquid state is likely highly entangled, and is \textit{not} a simple symmetry-breaking ground state.

Given that we always work on cylinders of finite extent, we cannot exclude the possibility that the peaks in Figs.~\ref{fig:SVN}(a,b) are due to surface critical phenomena \cite{binder1990critical,diehl1997theory} driven by a phase transition at the edge. Indeed, in Fig.~\ref{fig:Edge}(a), which shows a profile of the liquid regime on a wide cylinder at $\delta$\,$=$\,$3.50$, $R_b$\,$=$\,$1.95$, we notice that the edges seek to precipitate the most compatible density-wave order at these fairly large values of the detuning. Nonetheless, the bulk resists any such ordering tendencies and the central region of the system remains visibly uniform, with only slight perturbations from the open boundaries. In fact, the bulk fails to order \textit{despite} being at a detuning for which the system energetically favors a maximal (constrained) packing of Rydberg excitations, as is also evidenced by the nearby staggered and nematic phases above and below the liquid regime, respectively. It is perhaps worth noting that in one spatial dimension, the comparable regions lying between the different $\mathbb{Z}_n$-ordered states at large detuning are known to belong to a Luttinger liquid phase \cite{fendley2004competing}.

In order to eliminate end effects, it is often useful to first evaluate the ground-state energy per site for an infinitely long cylinder by subtracting the energies of finite cylinders of different lengths but with the same circumference \cite{yan2011spin,stoudenmire2012studying,PhysRevLett.110.127205,PhysRevLett.111.257201}. Such a subtraction scheme cancels the leading edge effects, leaving only the bulk energy of the larger system. In particular, this procedure enables us to quantify the influence of the boundaries on thermodynamic properties of the system such as the susceptibility. 
Using two cylinders of fixed width, an estimate of the bulk energy can be found by subtracting the energy of the smaller system from that of the larger. The (negative of the) second derivative of this quantity with respect to the detuning defines the \textit{bulk} susceptibility $\chi_b$---this gives us the susceptibility in the center of the cylinder with minimal edge effects. Figure~\ref{fig:Edge}(b) presents the variation of $\chi_b$ with detuning at $R_b$\,$=$\,$1.90$ for the YC6 family: we see that the local maximum of the susceptibility reported in Fig.~\ref{fig:SVN}(a) is still identifiable, but its precise location is shifted to slightly higher $\delta$. Analogously, we study the bulk susceptibility for wider YC8 cylinders and find, once again, a distinct peak corresponding to the onset of the liquid regime. Although this peak persists in a purely bulk observable, its magnitude is diminished: for example, the relative change between the local maximum and the minimum (shoulder) immediately adjacent to it on the right (left) differs by approximately a factor of four (ten) between $\chi_b$ and $\chi$ for the YC8 cylinder.
Hence, the behavior of the susceptibility could be indicative of an edge phase transition but whether this is accompanied by, or due to, a change in the bulk wavefunction is presently unclear.

\begin{figure*}[t]
\includegraphics[width=\linewidth]{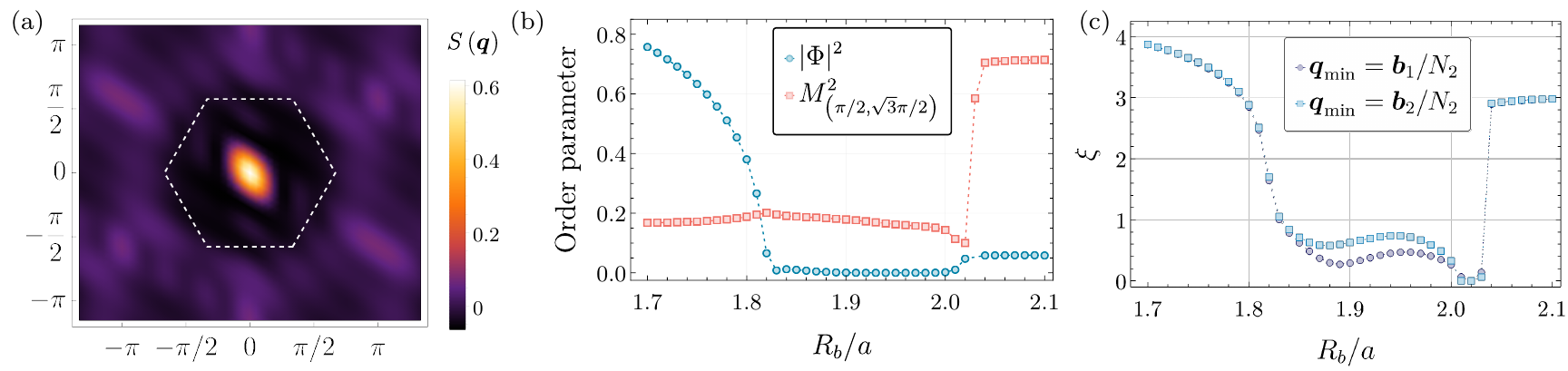}
\caption{\label{fig:DCut}\textbf{Properties of the liquid regime.} Here, we focus on the line $\delta = 3.3$ while varying the blockade radius. (a) The structure factor $S(\vect{q})$, at $R_b$\,$=$\,$1.95$, is featureless with no discernible ordering peaks. (b) The order parameters $\lvert \Phi \rvert^2$ and $M^2_{\vect{Q}}$ characterizing the nematic and staggered phases, respectively. Both develop a clear trough in the liquid region between the two phases, indicating the lack of symmetry-breaking order therein. (c) The correlation lengths calculated from \eqref{eq:SF}: for $1.9 \le R_b \le 2.0$, $\xi$ is smaller than one lattice spacing, so all correlations are short-ranged. }
\end{figure*}

Next, we investigate the properties of this liquid regime in more detail and demonstrate that---as preempted by its name---it does not possess any long-range density-wave order. This diagnosis of liquidity is best captured by the static structure factor. In stark contrast to the panels in Fig.~\ref{fig:SF}, $S(\vect{q})$ is featureless within the liquid regime [Fig.~\ref{fig:DCut}(a)] with the spectral weight distributed evenly around the extended Brillouin zone. 

This unordered nature is reflected in [Fig.~\ref{fig:DCut}(b)], where we plot the order parameters characterizing the surrounding symmetry-broken states. The order parameters defined earlier are found to be nonzero in both the nematic and staggered phases but are smaller by an order of magnitude in the liquid regime; this is compatible with a vanishing $\lvert \Phi \rvert^2$ and $M^2_{\vect{Q}}$ in the thermodynamic limit. In the process, we also find that the transition from the nematic (staggered) phase to the liquid regime appears to be second-order (first-order), which is consistent with the expectations for the QPT into a $\mathbb{Z}_2$ QSL in the dimer models, as we have discussed in the previous section. We do not observe any signatures of a phase transition \textit{within} the liquid regime.

Moreover, one can also define a correlation length from the structure factor as \cite{sandvik2010computational}
\begin{equation}
\xi (\vect{Q}, \vect{q}_\mathrm{min}^{}) = \frac{1}{\lvert \vect{q}_\mathrm{min}^{}\rvert} \sqrt{\frac{S(\vect{Q})}{S(\vect{Q}+ \vect{q}_\mathrm{min}^{})}-1},
\end{equation}
where $\vect{Q}+ \vect{q}_\mathrm{min}$ is the allowed wavevector immediately adjacent to the peak at $\vect{Q}$.
The correlation lengths obtained in the liquid are found to be smaller than the lattice constant, as plotted in Fig.~\ref{fig:DCut}(c), thus highlighting the lack of order. The qualitative behavior of $\xi$ is the same along both directions on the cylinder and mirrors that of the order parameter. On either side of the liquid region, the correlation lengths follow an increasing trend as long-range order develops deep in the solid phases. We have further verified that the bond-bond correlation functions  
\begin{equation}
C^{}_{(i,j),(k,l)} =4 [\langle(n^{}_i\cdot n^{}_j)(n^{}_k\cdot n^{}_l)\rangle-\langle n^{}_i\cdot n^{}_j \rangle\langle n^{}_k\cdot n^{}_l \rangle]
\end{equation}
are also short-ranged in the liquid regime.

So far, our numerics point to a gapped (see Fig.~S3), disordered candidate for the ground state of the liquid regime---these properties are all consistent with the behavior expected for a $\mathbb{Z}_2$ QSL, so it is natural to ask whether this region potentially harbors a topological phase. Although QSLs have long been fingerprinted by what the states are not i.e., by the \textit{absence} of ordering, more recently, it has been understood that the essential ingredient for a QSL is the \textit{presence} of massive quantum superposition leading to an anomalously high degree of entanglement \cite{savary2016quantum, grover2013entanglement}. Accordingly, we search for positive indications of a QSL in the liquid regime by calculating the topological entanglement entropy (TEE) \cite{kitaev2006topological, levin2006detecting} in Sec.~II of the SI. For a QSL phase, the value of the TEE is universal and positive, representing a constant reduction to the area law entropy. Importantly, the TEE arises entirely from nonlocal entanglement and is topological in origin. While we do find indications of an enhanced long-range entanglement entropy (Fig.~S4), this does not serve as conclusive evidence for a $\mathbb{Z}_2$ QSL as a finite $\gamma$\,$\sim$\,$\ln 2$ has also been documented for a valence bond solid in a different model \cite{PhysRevLett.110.127205}. Additionally, the TEE can suffer from strong finite-size effects on cylinders, leading to false signatures, and thus, cannot always reliably distinguish between different quantum phases \cite{gong2014plaquette}.

\section*{Discussion and outlook}

Based on numerical and theoretical analyses, we showed that the kagome lattice Rydberg atom array constitutes a promising platform for studying strongly correlated phenomena that supports not only a rich variety of quantum solids, but also, potentially,  a highly entangled liquid regime. We argued 
that the liquid region could host a state corresponding to an elusive phase with topological order using its
placement in the global phase diagram of triangular lattice quantum dimer models, and theories of their quantum phase transitions.
Our numerical study examined a number of signatures of the possible topological order and its associated phase transitions; although none of these computations conclusively confirm the existence of a topological phase for the available system sizes, they collectively point to interesting physics that merits further investigation.

This work can be extended in several directions. As DMRG is neither an unbiased method nor free from finite-size effects, it would be worthwhile to more completely quantify these uncertainties in future theoretical works, and definitively establish the nature of the liquid regime. 
A number of extensions to the present model can also be envisioned, e.g., by utilizing various atom arrangements as well as  multiple hyperfine sublevels or Rydberg atomic states to probe a variety of quantum entangled  phases.

Furthermore, we expect the phase diagram in Fig.~\ref{fig:PD}(e) to serve as a valuable guide to the detailed experimental studies of frustrated systems using Rydberg atom arrays. Specifically, both solid and liquid regimes can be reached starting from a trivial product ground state by adiabatically changing the laser detuning across the phase transitions, as was demonstrated previously \cite{bernien2017probing}. For experiments with $N$\,$\sim$\,$\mc{O}(10^2)$ Rb atoms coupled to a 70S Rydberg state, the typical Rabi frequencies involved can be up to $(2\pi)$\,$\times$10\,MHz. With these driving parameters, sweeps over the detuning range $0$\,$\leq$\,$\delta$\,$\lesssim$\,$5\,\Omega$, at interatomic spacings such that $R_b/a \lesssim 3.5$, have already been achieved in one-dimensional atom arrays \cite{endres2016atom, levine2018high, Omran570}. Hence, coherently preparing all the different many-body ground states and observing their fundamental characteristics should be within experimental reach in two-dimensional systems as well.

While the solid phases can be detected directly by evaluating the corresponding order parameter, the study of any possible QSL states in the liquid regime (or more generally, on Rydberg platforms) will require new approaches. 
%may prove to be less straightforward. With this caveat in mind, we outline here a few approaches that can be employed for the detection of a general $\mathbb{Z}_2$ QSL on Rydberg platforms. 
In particular, measuring the statistics of microscopic state occupations \cite{bernien2017probing} or the growth of correlations \cite{keesling2019quantum} across reversible QPTs could prove to be  informative. In a Rydberg liquid, one can think of creating and manipulating topologically stable excitations, which cannot disappear except by pairwise annihilation with a partner excitation of the same type. The excitation types would correspond to the three nontrivial anyons of the $\mathbb{Z}_2$ spin liquid, and each should manifest as a characteristic local (and stable) ``lump'' in the density of atoms in the excited Rydberg state; interference experiments between such excitations could be used to scrutinize braiding statistics.
The dynamic structure factor can also provide signatures of fractionalization: dispersive single-particle peaks will be observed in the disordered phase, while a two-particle continua would appear in a region with $\mathbb{Z}_2$ topological order. Detailed study of such spectra could yield the pattern of symmetry fractionalization \cite{Meng18,ZiYang2020dimers, becker2018diagnosing}. 
Other directions include more direct measurements of the topological entanglement entropy \cite{Islam:2015cm,Brydges:2019fp}.
Finally, classical and quantum machine learning techniques \cite{PhysRevB.96.195145,PhysRevB.96.245119,carrasquilla2017machine,cong2019quantum} could be useful for measuring  nonlocal topological order parameters associated with spin liquid states.

\matmethods{The DMRG calculations were performed using the ITensor Library \cite{ITensor}.  Further numerical details are presented in Sec.~I of the SI. 
%
%\subsection*{Subsection for Method}
%Example text for subsection.
}

\showmatmethods{} % Display the Materials and Methods section

\acknow{We acknowledge useful discussions with Subhro Bhattacharjee, Meng Cheng, Yin-Chen He, Roger Melko, Roderich Moessner,  William Witczak-Krempa, Ashvin Vishwanath, Norman Yao, Michael Zaletel, and especially the team of Dolev Bluvstein, Sepehr Ebadi, Harry Levine, Ahmed Omran, Alexander Keesling, and Giulia Semeghini. The authors are grateful to Marcus Bintz and Johannes Hauschild for pointing out the possibility of an edge transition and sharing their results.
We also thank Adrian E. Feiguin for benchmarking the ground-state energies observed in our DMRG calculations.
R.S. and S.S. were supported by the U.S.~Department of Energy under Grant $\mbox{DE-SC0019030}$. W.W.H., H.P. and M.D.L. were supported by the U.S.~Department of Energy under Grant $\mbox{DE-SC0021013}$,  the Harvard--MIT Center for Ultracold Atoms, the Office of Naval Research, and the Vannevar Bush Faculty Fellowship. W.W.H.~was additionally supported by the Gordon and Betty Moore Foundation's EPiQS Initiative, Grant No.~GBMF4306, and the NUS Development Grant AY2019/2020. The computations in this paper were run on the FASRC Cannon cluster supported by the FAS Division of Science Research Computing Group at Harvard University.}

\showacknow{} % Display the acknowledgments section

\noindent
{\em Note Added:} Another work which will appear in this arXiv posting studies the quantum phases of Rydberg atoms but in a different arrangement, where atoms occupy {\it links} of the kagome lattice \cite{Verresen20}.
% Bibliography
\bibliography{Ryd2D_Refs}



\section*{Supplementary material}

# Supporting Information for
# "Quantum phases of Rydberg atoms on a kagome lattice"

Rhine Samajdar,[1] Wen Wei Ho,[1, 2] Hannes Pichler,[3, 4] Mikhail D. Lukin,[1] and Subir Sachdev[1]

[1]*Department of Physics, Harvard University, Cambridge, MA 02138, USA*
[2]*Department of Physics, Stanford University, Stanford, CA 94305, USA*
[3]*Institute for Theoretical Physics, University of Innsbruck, Innsbruck A-6020, Austria*
[4]*Institute for Quantum Optics and Quantum Information, Austrian Academy of Sciences, Innsbruck A-6020, Austria*

The supplementary information presented here contains:

I. Technical details of DMRG computations, including a comparison of our results between different geometries that may or may not favor the solid phases breaking lattice symmetries.

II. Data on the apparent long-range entanglement entropy detected on cylinders of finite length.

III. Connection of the Rydberg Hamiltonian to previously studied models of hard-core bosons with ring-exchange interactions that are known to harbor a spin liquid phase.

IV. Exact diagonalization results on a 48-site torus confirming the existence of all the phases seen with DMRG and examining the excitation spectrum above the ground state.

## I. METHODS: DMRG IN TWO DIMENSIONS

Our numerical results in the main text are obtained from large-scale simulations of the Rydberg Hamiltonian using the density-matrix renormalization group (DMRG) [1–4], implemented with the ITensor library [5]. The remarkable success of DMRG is today understood to be attributable to an underlying matrix product structure, as the method operates on a particular class of quantum states [6–8] of the form

$$|\Psi\rangle = \sum_{\tau_1, \ldots, \tau_n} \sum_{b_1, \ldots, b_{n-1}} A_{b_1}^{\tau_1} A_{b_1 b_2}^{\tau_2} A_{b_2 b_3}^{\tau_3} \cdots A_{b_{n-1}}^{\tau_n} |\tau_1, \ldots, \tau_n\rangle,$$

where $A$ denotes matrices with physical indices $\tau$ and link indices $b$. The DMRG algorithm finds the optimal matrix product state (MPS) representation of the many-body ground state in this variational space of wavefunctions.

While originally formulated as a tool for studying strongly correlated one-dimensional quantum systems, DMRG can be extended to two dimensions by mapping the 2D lattice on to a 1D chain with longer-range interactions. Since open boundaries act as effective pinning fields [9, 10] for the Rydberg excitations, the bulk properties of the model are (ideally) best studied on a torus. However, the imposition of fully periodic boundary conditions requires squaring the number of states needed for a given accuracy [11], so we instead place the system on a cylinder, with open boundaries along $\mathbf{a}_1$ but periodic ones along $\mathbf{a}_2$.

### A. Lattice geometry

To avoid spurious effects due to sharp edges, following Refs. [12–14], we work with a geometry such that each unit cell at the right boundary of the cylinder contains only two sites. The resultant lattice, labeled as YC ($2N_2$), has a noninteger number of unit cells, with a total of $N_2 \times (3N_1 + 2)$ sites. Such a geometry has the added advantage of stabilizing the various ordered states. This implies that within the liquid regime detected in Fig. 1(e), the proximate solid orders are unstable *despite* being explicitly favored, which constitutes further evidence against an ordered ground state.

In Fig. S1, we demonstrate that our results are qualitatively the same on lattices with and without an integer number of unit cells, and that a featureless liquid state exists in both cases. The specific density-wave profiles in the solid phases, however, may be sensitive to the jagged edges and can differ from those illustrated in the main text [Figs. 1(b–d)], as typified by the nematic order in Fig. S1(a): this is because the boundaries seed two distinct symmetry-broken configurations from either end, which necessarily merge in the center of the system, forming a domain wall.

### B. Convergence

For the DMRG calculations, our protocol entails first carrying out a large number of sweeps at relatively small bond dimensions ($d \sim 100$) before increasing $d$ progressively at later stages. At each diagonalization step, we allow for up to six iterations of the Davidson algorithm to facilitate proper convergence. To assist the build-up of long-range correlations, it is useful to initially add a small "noise" term [15] to the density matrix, which is turned off in subsequent sweeps. Recognizing that DMRG is a variational algorithm [16], we specifically ensure that the calculation is not stuck in a metastable state [17, 18] by checking for convergence with respect to both bond dimension and number of sweeps. For instance, in Fig. S2, we show the behavior of ground-state properties as a function of $d$ for the largest of the cylinders considered

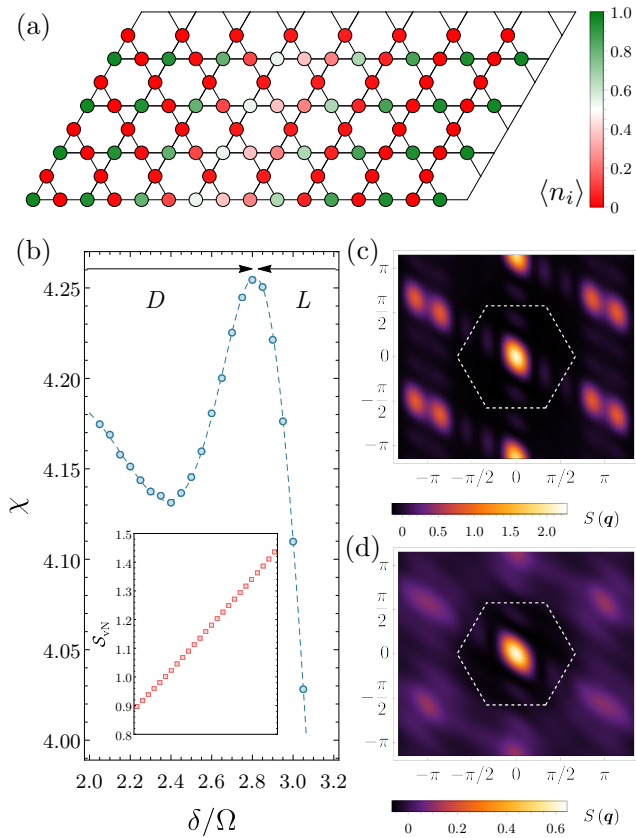

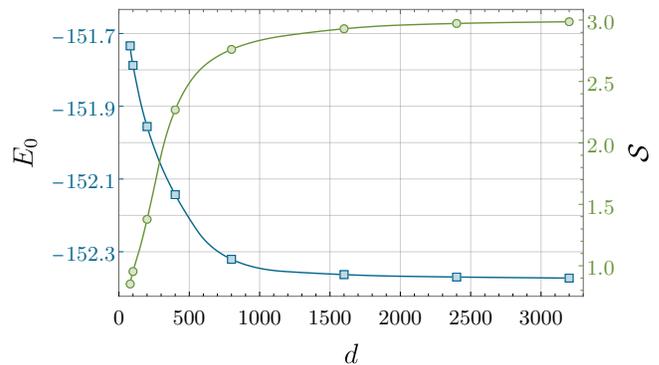

FIG. S2. **Convergence of DMRG calculations.** For the YC12 geometry, plotted here are the ground-state energy ($E_0$, blue) and the entanglement entropy on the central bond ($S$, green) as a function of the bond dimension $d$ at a generic point in the liquid regime ($\delta = 3.50$, $R_b = 1.95$). The system can be regarded to have converged to its true ground state, as evidenced by the change of $< 0.002\%$ in $E_0$ upon increasing $d$ from 2400 to 3200.

FIG. S1. **Ground states on kagome lattices with an integer number of unit cells.** (a) Nematic ordered state on a $8 \times 4$ lattice at $\delta = 3.5$, $R_b = 1.7$, formed by a superposition of two competing $C_3$-symmetry-broken states nucleated by the edges. (b) The susceptibility, at $R_b = 1.9$, exhibits a distinct maximum occurring at the same position as for the YC8 cylinders in Fig. 5(a). $D$ and $L$ indicate the trivial disordered and correlated liquid regimes, respectively. The von Neumann entanglement entropy (inset) monotonically increases over the same range, in exact correspondence with Fig. 5(c). Static structure factors at $\delta = 3.5$ for (c) $R_b = 1.7$, and (d) $R_b = 1.9$; $S(\boldsymbol{q})$ shows conspicuous ordering peaks for the former, but is featureless for the latter.

in this work (YC12), at a representative point in the liquid regime ($\delta = 3.50$, $R_b = 1.95$): the satisfactory convergence confirms that we are able to numerically obtain accurate ground-state wavefunctions even for these fairly wide systems. The maximum truncation errors in the liquid region were found to be $\mathcal{O}(10^{-6})$.

## C. Energy gaps

Along with the ground state $|\psi_0\rangle$, a protocol similar to that described above can be used to also find the excited states, although it is computationally more expensive to do so. We can target the first excited state using the Hamiltonian $H' = H_{\mathrm{Ryd}} + wP_0$, where $P_0 = |\psi_0\rangle\langle\psi_0|$ is a projection operator and $w$ is an energetic penalty. Such

a computation leads to Fig. S3, where we notice that the energy gap to the first-excited state found with DMRG, $\Delta$, is nonzero in the liquid region and its magnitude is consistent with the values obtained from exact diagonalization calculations on a torus with the same width (see Sec. IV).

Let us outline here the general expectations for the gap in a $\mathbb{Z}_2$ quantum spin liquid (QSL) phase. As opposed to the fourfold degeneracy on a torus, a $\mathbb{Z}_2$ QSL would only be twofold degenerate on a cylindrical system—this is because the ground-state energy in two of the sectors involves an additional cost due to the presence of a pair of anyons localized at the ends of the cylinder [19]. However, DMRG preferentially converges to *one* of the quasidegenerate ground states and, in particular, to a minimally entangled state (MES). This well-recognized entanglement barrier between different MESs [20] accounts for the apparent absence of a topological degeneracy in several numerical studies [10, 21–24]. The splitting between the two quasidegenerate ground states (on a cylinder) belonging to different topological sectors scales as $\sim L_h\, e^{-cL_v}$, and is likely to be orders-of-magnitude smaller than $\Delta$ [23]. For the Rydberg model, it remains to be seen whether all the four topological sectors of a $\mathbb{Z}_2$ QSL can be constructed in the liquid regime by threading anyon lines through an infinite cylinder [25].

Note that in the ordered phases, we see that the gap is nearly zero, which is a numerical indicator of the ground state being degenerate [23]. Thus, the observation of a finite gap in the intermediate region rules out *any* type of symmetry-breaking order there, including all possibilities not considered explicitly via the correlation functions [23]. However, a nonzero $\Delta$ does not distinguish between a trivial disordered and a potential topological liquid phase, and could also arise from edge excitations.

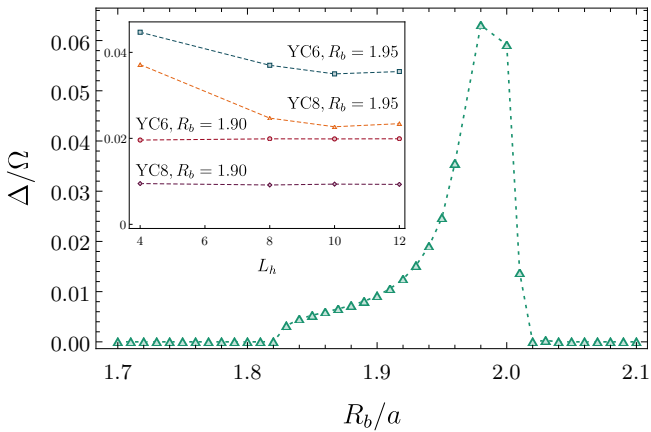

FIG. S3. **Energy gap to the first-excited state detected with DMRG.** Along the line $\delta = 3.3$, $\Delta$ is nonzero over the intermediate range of $R_b$ corresponding to the liquid regime between the nematic (left) and staggered (right) orders. Inset: the gap remains nonvanishing on several cylinders of different circumferences and lengths, so the liquid is demonstrably gapped in the thermodynamic limit.

## II. EXPLORING TOPOLOGICAL ORDER

For a gapped system in 2D, assuming that the boundary, of length $L$, between two subsystems is smooth (i.e., devoid of corners), the corresponding von Neumann entanglement entropy obeys an "area" law with a potential constant subleading correction

$$\mathcal{S}_{\rm vN} = \alpha L - \gamma + \mathcal{O}\left(1/L\right), \tag{S1}$$

where $\alpha$ is a model-specific nonuniversal coefficient, and $\gamma$ is the topological entanglement entropy (TEE). In a phase without topological order, $\gamma$ is trivially zero. For a QSL phase, however, the value of $\gamma$ is universal and positive, representing a constant reduction to the area law entropy. The TEE arises entirely from nonlocal entanglement and is topological in origin: in fact, it is known that $\gamma = \ln D$ with $D = \sqrt{\sum_\lambda d_\lambda^2}$, where $d_\lambda \geq 1$ is the quantum dimension of the quasiparticle $\lambda$ [26]. Hence, $\gamma$ probes the anyon content of the topological order, so $D > 1$ ($\gamma > 0$) naturally implies that the state supports fractionalized excitations.

Following the prescription proposed by Jiang *et al.* [22, 23], one can read off $\gamma$ by placing the system on infinite cylinders with varying circumferences $L_v \equiv 2N_2$, and extrapolating the EE [27] per the scaling form of (S1). In practice, this procedure is efficient whenever all correlation lengths are much shorter than the width of the cylinder [28]—as is the case in Fig. 7(b). As pointed out by Refs. [22, 23], a nonzero $\gamma$ can potentially arise from two distinct sources: a symmetry-breaking contribution that enhances the total EE, and a topological piece which reduces it. However, the former, which arises from global entanglement of the entire system, can be eliminated by increasing the length $L_h$ of the system at fixed width

$L_v$ so that the DMRG algorithm converges to an MES amongst the manifold of ground states that are degenerate in the limit of infinite system size [29]. On the kagome lattice, the infinite-cylinder limit ($L_h = \infty$) is believed to be well approximated when $L_h > L_v$ [30]. Using the thus obtained entropies, we compute the TEE for different points along the line $R_b = 1.95$. At $\delta = 0.5$, where the system is in the trivial disordered phase, we find $\gamma \approx 0$ [see inset of Fig. S4(a)], so the area law is strictly obeyed, as is expected for a state without long-range entanglement that can be smoothly deformed into a product state. Contrarily, at $\delta = 3.5$ (in the middle of the liquid regime), the best linear fit of the EE gives $\gamma = 0.64$. Although this value is comparable to the theoretically known TEE of $\ln 2$ for a $\mathbb{Z}_2$ spin liquid [31, 32], it should not be interpreted as firm evidence for a QSL phase. This is because strong finite-size effects on cylinders [33] are known to often produce spurious contributions to the TEE since the "replica" length scale over which $\gamma$ converges can be arbitrarily larger than the physical two-point correlation length [28].

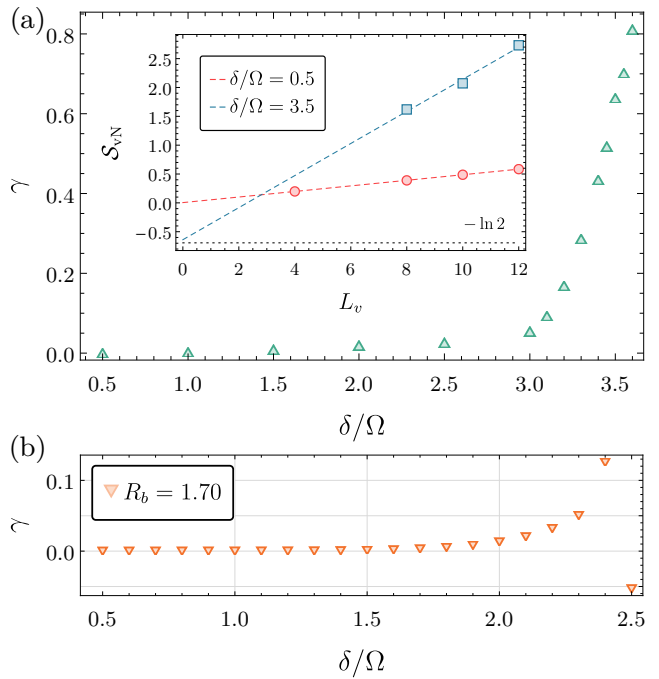

FIG. S4. **Subleading corrections to the area law in the liquid regime.** (a) Plotting the variation of the TEE with $\delta$ at fixed $R_b = 1.95$, we see that $\gamma$ vanishes in the disordered phase and starts to deviate from zero in the vicinity of the previously determined susceptibility maximum. Inset: linear extrapolation of the EE with the cylinders' circumferences according to (S1), demonstrating a nonzero and negative intercept in the liquid regime. (b) The long-range entanglement entropy grows with increasing $\delta$ along $R_b = 1.70$, and changes sign across the QPT into the nematic phase beyond $\delta = 2.40$. Its overall magnitude, however, is much smaller than the TEE in part (a).

Beyond the precise value of $\gamma$—which is known to be sensitive to numerical details [34] and challenging to estimate accurately [33, 35–38]—we emphasize the positive sign of the TEE as any discrete-symmetry-breaking state, should yield a constant correction to the EE of *opposite* sign [23]. We verify the robustness of this observation by plotting $\gamma$ as a function of $\delta$, for $R_b = 1.95$, in Fig. S4(a). The TEE remains zero for an extended range of $\delta$ and only starts to rise in a crossover region centered around the susceptibility maximum at $\delta = 3.2$, calculated in Fig. 5(a)[39]. The sign of $\gamma$, and more importantly, the coincidence of the apparent onset of a nonzero TEE with the peaks in $\chi$ and $\mathcal{F}$ determined earlier, are together suggestive of a possible transition into a topological QSL state.

Although $\gamma$ does not seem to saturate to a constant value of $\ln 2$ in Fig. S4(a), such variations in the numerical TEE, depending on the precise point studied in parameter space, have been previously reported in the literature for other spin-liquid candidates [33, 37] as well. In our case, this effect could stem from the proximity to the string phase, which sets in at $\delta \sim 3.7$ on the YC12 cylinder. In the neighborhood of a second-order QCP in two dimensions, the EE is believed to possess subleading corrections to the area law [40]: $\gamma$, as defined by Eq. (S1), is then a nonzero universal constant, which we refer to as the long-range entanglement entropy (LREE) since it is not of topological origin. The LREE is difficult to determine analytically and the exact value of this

geometric constant is not known even for the Ising QCP. To estimate the contribution of the LREE to the TEE in Fig. S4(a), we compute $\gamma$ for variable $\delta$ at $R_b = 1.7$ as the system transitions into the nematic phase [Fig. S4(b)]. We find that the resultant magnitudes are not large enough to account for a TEE of $\mathcal{O}(\ln 2)$, leading us to conjecture that the net long-range entanglement entropy seen earlier in the liquid regime is not merely an artifact of the QCP alone. Admittedly, we cannot unambiguously rule out a scenario where there is a single direct QPT from the disordered phase to the string-ordered crystal, but we note the absence of any signal of a diverging correlation length in our finite-cylinder numerics, which would be expected in the same regime as the enhancement of the LREE if that were the case.

## III. BOSON MODELS WITH RING EXCHANGE

In this section, we demonstrate that the Rydberg Hamiltonian on the kagome lattice can be related, at least perturbatively, to certain well-studied models of hard-core bosons with "ring-exchange" interactions [41–43], which have previously been identified to host a QSL phase.

The mapping to hard-core bosons proceeds straightforwardly by associating each atom in the Rydberg (ground) state with the presence (absence) of a boson [44] on the corresponding lattice site. In the bosonic language, $H_{\text{Ryd}}$ can be reformulated as

$$H_{\text{Ryd}} \equiv H_0 + H_1; \qquad H_0 = V_1 \sum_{\langle i,j \rangle} n_i n_j + V_2 \sum_{\langle\langle i,j \rangle\rangle} n_i n_j + V_3 \sum_{\langle\langle\langle i,j \rangle\rangle\rangle} n_i n_j - \delta \sum_i n_i, \quad H_1 = \frac{\Omega}{2} \sum_i (b_i^\dagger + b_i), \quad \text{(S2)}$$

where $b_i^\dagger (b_j)$ is the boson creation (annihilation) operator, $n_i = b_i^\dagger b_i$ is the number operator, and $V_i$ stands for the repulsion strength between $i$-th nearest-neighbors with $V_1 = 27$ $V_2 = 64$ $V_3$ owing to the van-der-Waals nature of the interaction. While this Hamiltonian does not conserve the total number of bosons, we first derive an effective Hamiltonian that recovers the global U(1) symmetry broken by the $(b_i^\dagger + b_i)$ terms in Eq. (S2). This is motivated by considering the limit of large positive detuning such that boson number is effectively conserved. Using the symbols $\alpha, \beta$ to label sectors with a fixed number of bosons, and $m, n$ to denote states within each group, the matrix elements of the effective Hamiltonian are given by [45]

$$\langle m, \alpha | H_{\text{eff}} | n, \alpha \rangle = E_{m,\alpha} \delta_{m,n} + \langle m, \alpha | H_1 | n, \alpha \rangle + \sum_{l, \beta \neq \alpha} \frac{\langle m, \alpha | H_1 | l, \beta \rangle \langle l, \beta | H_1 | n, \alpha \rangle}{2} \left( \frac{1}{E_{m,\alpha} - E_{l,\beta}} + \frac{1}{E_{n,\alpha} - E_{l,\beta}} \right),$$
$$\text{(S3)}$$

where $E$ is the (purely classical) energy of a given configuration as determined by $H_0$ alone. Let us now evaluate Eq. (S3) term by term. Consider a second-order hopping process where an existing boson on a given site, say $i'$, is annihilated first, followed by the creation of a boson

on an adjacent site $j'$. Crucially, owing to the Rydberg blockade, the hopping amplitude will be severely reduced if any of the three nearest-neighboring sites of $j'$ (besides $i'$) are occupied. Treating this effect probabilistically, we replace the energy denominators in Eq. (S3) by their

configurational averages

$$\left\langle \frac{1}{E_{m,\alpha} - E_{l,\beta}} \right\rangle = \left\langle \frac{1}{E_{n,\alpha} - E_{l,\beta}} \right\rangle$$
$$= \frac{1}{-\delta} \mathcal{P}^3 + \frac{3}{V_1 - \delta} \mathcal{P}^2 (1 - \mathcal{P}), \quad \text{(S4)}$$

neglecting terms with $p\,V_1$ ($p > 1$) in their denominators. Here, $\mathcal{P}$ is the probability of finding an unoccupied site; recognizing that the Rydberg liquid appears in proximity to a phase with a filling fraction of one-third, we set $\mathcal{P} = 2/3$. With this assumption, the leading-order matrix elements of $H_{\text{eff}}$—from the hopping described above—are given by

$$\frac{(\Omega/2)^2}{2} \left( \frac{1}{-\delta} + \frac{1}{-\delta} \right) \mathcal{P}^3. \quad \text{(S5)}$$

Next, we consider the reverse process in which a boson is first created on a given site and then an existing boson is annihilated on a neighboring site. Likewise, the approximate matrix elements of $H_{\text{eff}}$ in this case are

$$\frac{(\Omega/2)^2}{2} \left( \frac{1}{\delta - V_1} + \frac{1}{\delta - V_1} \right) \mathcal{P}^3. \quad \text{(S6)}$$

Note that Eq. (S6) is already of order $1/V_1$ and can be neglected in comparison to Eq. (S5) since $V_1 \gg \delta$.

Naively, the analysis above suggests that there are processes by which a particle can hop arbitrary distances, but these cancel between the contributions of the $N - 1$- and $N + 1$-boson subspaces [45] at this order, and the only surviving hopping terms connects adjacent sites. This leads us to the effective Rydberg Hamiltonian

$$H_{\text{eff}} = -\sum_{\langle i,j \rangle} t \left( b_i^\dagger b_j + \text{H.c.} \right) + V_1 \sum_{\langle i,j \rangle} n_i n_j + V_2 \sum_{\langle\langle i,j \rangle\rangle} n_i n_j + V_3 \sum_{\langle\langle\langle i,j \rangle\rangle\rangle} n_i n_j - \delta \sum_i n_i, \quad \text{(S7)}$$

with $t = \Omega^2 \mathcal{P}^3 / (4\,\delta)$. The reason behind this formal manipulation is that it allows us to rewrite

$$H_{\text{eff}} = -t \sum_{\langle i,j \rangle} \left( b_i^\dagger b_j + \text{H.c.} \right) + 2\mathcal{V} \left( \sum_{\langle i,j \rangle} n_i n_j + \sum_{\langle\langle i,j \rangle\rangle} n_i n_j + \sum_{\langle\langle\langle i,j \rangle\rangle\rangle} n_i n_j \right) - \delta \sum_i n_i + \mathcal{H}_{\text{def}} \quad \text{(S8)}$$

$$= -t \sum_{\langle i,j \rangle} \left( b_i^\dagger b_j + \text{H.c.} \right) + \mathcal{V} \sum_{\bigcirc} \left[ \left( n_{\bigcirc} - \frac{\mu}{4\mathcal{V}} \right)^2 - \frac{\mu^2}{16\mathcal{V}^2} \right] + \mathcal{H}_{\text{def}} \equiv \mathcal{H}_{\text{b}} + \mathcal{H}_{\text{def}}, \quad \text{(S9)}$$

where $\mathcal{H}_{\text{b}}$ only includes homogeneous interactions while $\mathcal{H}_{\text{def}}$ can be viewed as a deformation thereof that encompasses all the distance-dependent nonuniformity in $H_{\text{eff}}$. In Eq. (S9), $n_{\bigcirc}$ is the number of particles in each of the hexagons of the kagome lattice, $\mu = \delta + 2\mathcal{V}$ is the effective chemical potential, and $\mathcal{V}$ is a single short-range repulsion strength that we will specify shortly. It is easy to see that for $\mu = (4, 8, 12)\mathcal{V}$, the second term of Eq. (S9) is minimized by having $(1, 2, 3)$ bosons per hexagon respectively or equivalently, a filling fraction of $f = (1/6, 1/3, 1/2)$. The undeformed model $\mathcal{H}_{\text{b}}$, at half-filling, is known to exhibit a superfluid-insulator transition at $(\mathcal{V}/t)_c \approx 19.8$, and the insulating phase is a topologically ordered $\mathbb{Z}_2$ Mott insulator [42]. However, at both $1/3$ and $1/6$ fillings, the model also has a $\mathbb{Z}_2$ spin liquid regime as shown by Ref. [43] following a mapping onto the triangular-lattice quantum dimer model. As the ratio $\mathcal{V}/t$ has to *exceed* a certain critical value to obtain the QSL phase, one should compare $\mathcal{V}$ in $H_{\text{eff}}$ to the smallest interaction scale in Eq. (S7); accordingly, we identify $2\mathcal{V} = V_3$. Supplementing this equation with the relation $\delta = 6\mathcal{V}$ and the derived expression for $t$, one can easily solve for $\{\delta/\Omega, V_3\}$. Roychowdhury *et al.* [43]

showed that the parameter ranges realizing the spin liquids at $1/2$ and $1/3$ filling are nearly identical, which enables us to use the previously stated estimate of $(\mathcal{V}/t)_c$ by Isakov *et al.* [42] in our calculation. Taking, for instance, $\mathcal{V}/t = 20 \gtrsim (\mathcal{V}/t)_c$, we find $R_b/a = 1.997$ and $\delta/\Omega = 2.981$, which is reasonably close to the region observed numerically for the Rydberg liquid regime.

## IV. EXACT DIAGONALIZATION STUDIES

In this section, we supplement the DMRG simulations of the main text with exact diagonalization (ED) studies of the kagome lattice Rydberg Hamiltonian. While ED techniques are restricted to system sizes smaller than those accessible with DMRG, they are completely unbiased and offer a complementary viewpoint as one is able to probe features that are harder to extract using DMRG, such as spectral gaps to higher excited states as well as the full distribution of the ground-state wavefunction over the computational basis states. Furthermore, with ED, one has the ability to impose arbitrary boundary conditions such as toroidal ones, which help circumvent

edge effects but are challenging to handle with tensor network methods. We find that the ED numerics confirm the existence of at least four phases (see Fig. S6), as well as the natures of the ordered phases. However, we do not exactly observe the fourfold near-degeneracy of the ground state that one would expect for a topological $\mathbb{Z}_2$ liquid—this is not surprising given the large finite-size effects known to affect ED studies of spin systems on the kagome lattice [46, 47].

We consider here a 48-site cluster of linear dimensions $N_1 = N_2 = 4$ with fully periodic boundary conditions such that an atom located at position $\boldsymbol{r}$ is identified with those at positions $\boldsymbol{r} + N_\mu \boldsymbol{a}_\mu$ ($\mu = 1, 2$). The full Hilbert space is of (a rather intractable) dimensionality $2^{48} \approx 2.8 \times 10^{14}$, so we instead operate in the so-called "Rydberg-blockaded" space, where no two neighboring atoms on the lattice are allowed to be simultaneously excited. This leads to an effective Hamiltonian

$$H_{\text{eff}} = \sum_i \Omega \prod_{\langle i,j \rangle} \left(1 - n_j\right) S_i^x - \delta \sum_i n_i + \sum_{a < ||\boldsymbol{x}_i - \boldsymbol{x}_j|| \leq 2a} V_{ij} n_i n_j, \tag{S10}$$

where $S_i^x = (|r\rangle_i \langle g| + \text{H.c.})/2$, and $V_{ij} \equiv V\left(||\boldsymbol{x}_i - \boldsymbol{x}_j||/a\right)$; hereafter, we will set $\Omega = a = 1$ as before. The first term describes a spin-flip in the blockaded space, and the relation $\langle i, j \rangle$ in the projector specifies that sites $i$ and $j$ are nearest neighbors (NNs). In the last term, we sum over pairwise interactions of Rydberg atoms with mutual distances corresponding to second- and third-NNs; the first-NN repulsion strength is formally infinite due to the hard blockade constraint. The distance between any two sites is taken to be the shortest one on the torus. For $R_b \gtrsim 1$, the effective model (S10) captures the essential physics of the Rydberg Hamiltonian as NN interactions in the latter are so strong that there is an enormous energetic penalty for the simultaneous excitation of two neighboring atoms. In particular, as the liquid regime occurs at $R_b \sim 1.9$, the effective model should presumably bring out its existence as well as its universal properties.

It is useful to note that the Hamiltonian $H_{\text{eff}}$ is also invariant upon translations in the $\boldsymbol{a}_1$ and $\boldsymbol{a}_2$ directions, and the spectrum can therefore be decomposed into momentum sectors. For concreteness, let $T_1$ and $T_2$ represent the operators implementing such shifts in the respective directions. Given that $T_\mu^{N_\mu} = 1$ and $N_\mu = 4$, the eigenvalues of $T_\mu = e^{ik_\mu}$ range across $k_\mu = 2\pi n_\mu/4$ with $n_\mu = 1, 2, 3, 4$. It turns out that there are four sectors whose spectra are not unitarily related to one another: these are labeled by $(k_1, k_2) = (\pi/2, \pi/2), (\pi/2, \pi), (\pi, \pi), (2\pi, 2\pi)$, and have Hilbert space dimensions of 7587799, 7587792, 7590567, and 7590689, respectively. In total, six sectors are equivalent to $(\pi/2, \pi/2)$, six others to $(\pi/2, \pi)$, three to $(\pi, \pi)$, and one to $(2\pi, 2\pi)$. This momentum resolution is crucial for our ability to numerically treat a system of 48 spins with ED.

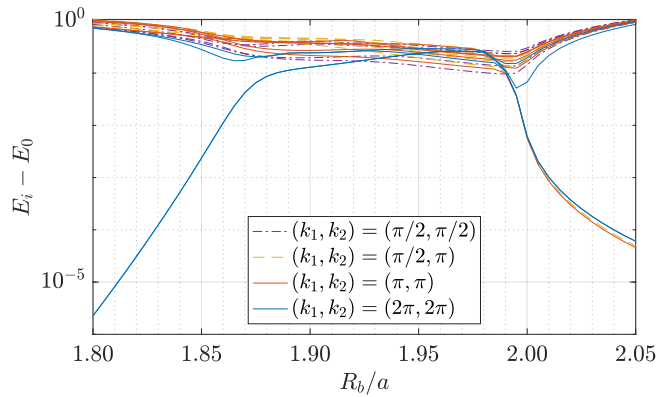

FIG. S5. **Spectral gaps of low-lying eigenstates.** Using ED, we trace the evolution of the gaps $E_i - E_0$ (in units of $\Omega$) for the first few states, with energies $E_i$, of the effective Hamiltonian (S10) as a function of $R_b$ at fixed $\delta = 3.8$ on a 48-site torus; the ground-state energy is $E_0$. Note that the lowest blue curve is twofold degenerate. The region $R_b < 1.85$ corresponds to the nematic phase with threefold ground state degeneracy, whereas the staggered phase prevails for $R_b > 2.0$ with twelvefold degeneracy. In between, we clearly see the appearance of a distinct phase.

In Fig. S5, we display the low-lying gaps $E_i - E_0$ between eigenstates with energy $E_i$ and the ground state, which has energy $E_0$, working at a fixed detuning, $\delta = 3.8$. For $R_b < 1.85$, there are two eigenstates with minuscule energy differences ($\sim 10^{-6}$ in units of $\Omega$) above the ground state. This implies that we should consider the first three states of the system as belonging to a "ground-state manifold". Indeed, it is easy to check that the decomposition of the ground-state wavefunction in terms of the computational basis states in the $z$-direction at $R_b = 1.8$ is dominated by the three configurations of nematic ordering. On the other hand, for larger $R_b > 2.0$, the system, once again, possesses many low-lying eigenstates with small energy splittings ($\sim 10^{-5}$–$10^{-4}\,\Omega$). In this case, we predominantly find the classical configurations corresponding to staggered order in the wavefunction decomposition, in alignment with the predictions from DMRG. However, for intermediate values of the blockade radius, where $1.9 \lesssim R_b \lesssim 2.0$, we notice a markedly different spectra, clearly showing the presence of an intervening phase.

Prompted by the observations above, we now investigate the various phases realized on the 48-site kagome lattice in further detail. In parallel with the main text, we first plot, for three different values of the blockade radius, the susceptibility $\chi = -\partial^2 E_0/\partial \delta^2$, which can serve as a convenient probe to hunt for quantum phase transitions (QPTs). In the thermodynamic limit, the susceptibility diverges at a quantum critical point but on a finite lattice, this divergence is inevitably rounded off and manifests itself as a local maximum of $\chi$. Focusing on this diagnostic, in Fig. S6(b), we see a peak occurring

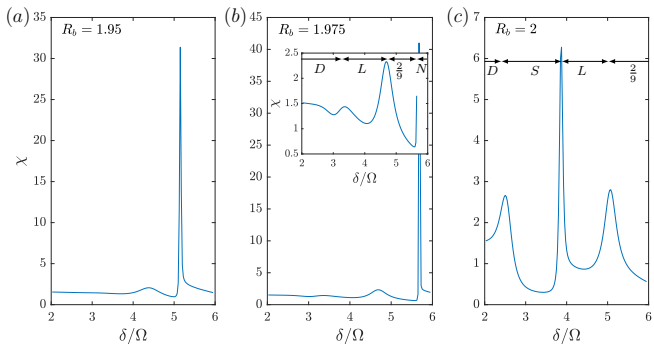

FIG. S6. **Susceptibilities observed in ED simulations**. The variation of $\chi = -\partial^2 E_0/\partial \delta^2$ with the detuning $\delta$ at a constant $R_b =$ (a) 1.95, (b) 1.975, and (c) 2.00, brings out several noticeable local maxima, which could be indicative of quantum phase transitions. The shorthand labels $N$, $S$, and 2/9 denote the extents of the nematic, staggered, and string phases, respectively. The susceptibility betokens that the QPT between the staggered phase and the liquid regime is first-order, in agreement with DMRG.

around $\delta \sim 3.2$ at $R_b = 1.975$. Since the phase at small $\delta$ is expected to be trivially disordered, we infer, albeit indirectly, that the region on the other side of the peak is possibly the correlated liquid regime—we will verify this hypothesis shortly. We emphasize, however, that a scaling analysis of the height of the peak with system size is necessary to confirm whether this signal is due to a genuine QPT. At even larger $\delta$, the system undergoes two successive transitions into first the string phase and thereafter, the nematic.

Figure S7 collates the momentum-resolved energy spectrum, the ground-state decomposition, and the structure factor in each of the four regions identified from the susceptibility. Inspecting the ground-state wavefunction at $\delta = 3.8$ [Fig. S7(b)] reveals that the distribution over computational states is rather spread out, with no particular configuration dominating the decomposition, in distinction to the discernible peaks for the string ($\delta = 5.25$) and nematic ($\delta = 6.00$) phases seen in Figs. S7(c) and (d), respectively. One can also generate representative classical configurations constituting the ground-state wavefunctions by sampling these distributions: from such snapshots, we find that the microstates in the liquid regime do not bespeak any particular order, whereas for the string and nematic phases, the ideal density-wave ordering is readily visible in local patches. This liquidity is also reflected in the static structure factor, which is mostly uniform in Fig. S7(b) but develops prominent features for higher values of $\delta$. Taken together, these three pieces of information indicate that the liquid state is *not* ordered. Finally, let us mention that the dimer-dimer correlator [Eq. (6)] is zero by construction (since we work in the blockaded space), thereby ruling out any valence bond crystal phases.

The energy spectra arrayed above present another in-

dependent method to potentially distinguish between the disordered and liquid regimes: for instance, in the zero-momentum sector, the first-excited state is always doubly degenerate in the former but unique in the latter. For a trivial paramagnet, the lowest-energy translationally invariant excitation should be a superposition of spin-flips, which is a property that holds throughout the phase. Therefore, the change in the character of the excitations between Figs. S7(a) and (b), conveyed by the differing $\boldsymbol{k} = 0$ spectra, suggests that the two regimes could be of different natures. Another nontrivial distinction pertains to which six states in the $(k_1, k_2)$ sectors constitute the lowest-lying ones (above the ground state at the $\Gamma$ point): in the disordered phase, these are the states in the $(k_1, k_2) = (0, \pi)$, and equivalent, sectors but, for the liquid, they belong to the $(k_1, k_2) = (0, \pi/2)$ and associated sectors.

Similar considerations apply for the $R_b = 2.0$ line scrutinized in Fig. S8, in exact analogy to Fig. S7. In the liquid regime [Fig. S8(c)], we notice that the energy splitting between the absolute ground state and the first few low-lying eigenstates, $E_i - E_0$, is very small at $\sim 0.05\,\Omega$. In the thermodynamic limit, three of these excited states could, in principle, have energies that approach that of the ground state with the rest remaining gapped, thereby forming the expected ground-state manifold of a $\mathbb{Z}_2$ spin liquid wherein each of the four states corresponds to an anyon type in the theory. While we are unable to conclusively detect whether this scenario occurs based on our ED simulations, we note that such a drawback is also present for several other ED studies [46–51] of kagome systems such as the spin-1/2 Heisenberg antiferromagnet [52] (which is believed to host a spin liquid phase), where it has been attributed to nontrivial finite-size effects.

In summary, our ED calculations provide a comprehensive picture of the low-energy physics of the kagome lattice Rydberg atom array. Our numerics on a 48-site cluster are broadly consistent with the results reported by DMRG, thus lending support to and strengthening the conclusions of the main text.

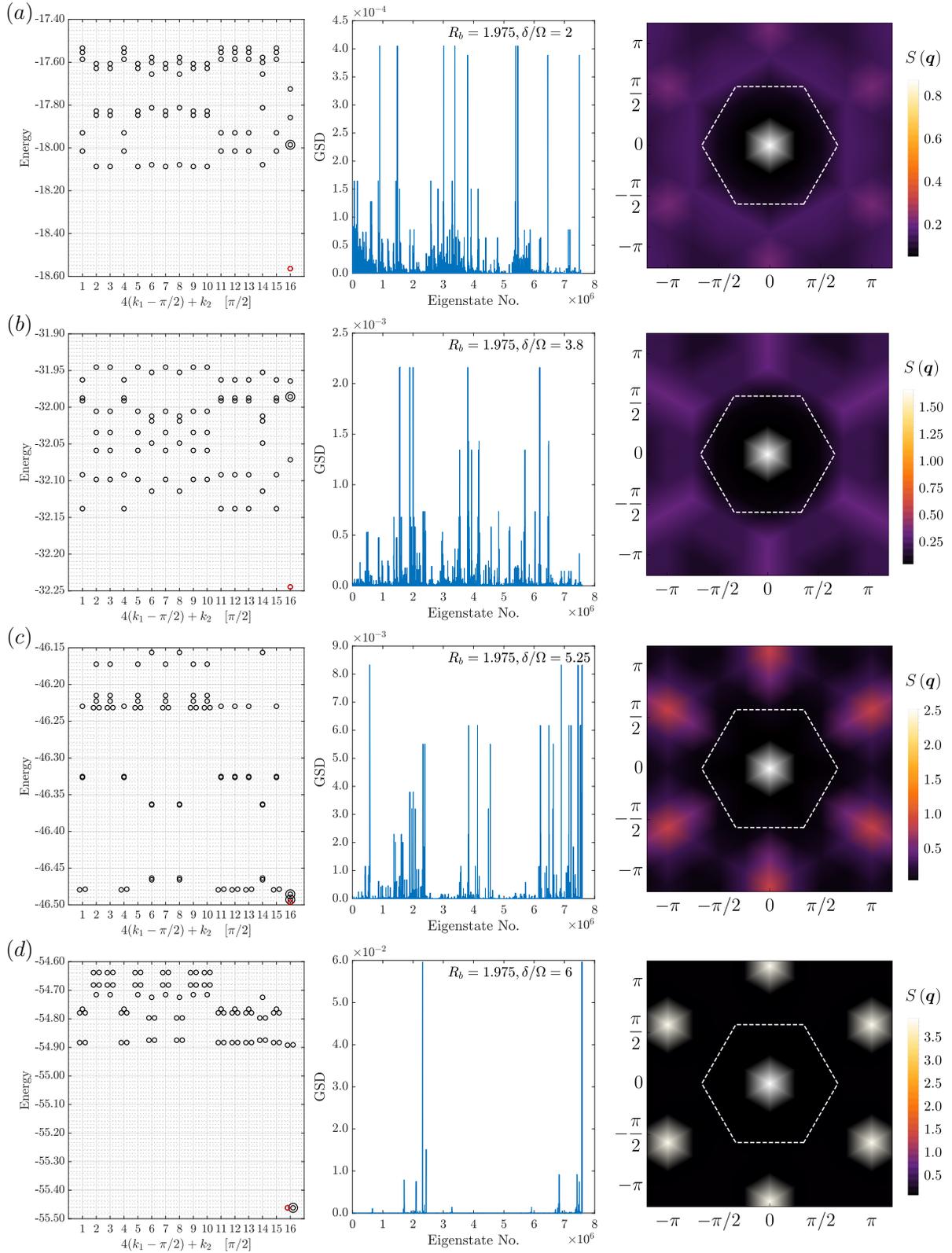

FIG. S7. **Excitation spectra and ground-state properties at** $R_b = 1.975$. Three of the chosen points in parameter space belong to the (a) disordered, (c) string, and (d) nematic phase, whereas (b) lies within the liquid regime. The leftmost column shows the low-lying spectrum. If two states are degenerate within machine precision, they are plotted as concentric circles; for the sake of visual resolution, if two levels are spaced less than $10^{-4}$ apart, they are depicted as being split horizontally. The central column illustrates the ground-state decomposition (GSD) in the zero-momentum sector of the state circled in red in the left panels—the blue bars represent the probability for each classical configuration, indexed along the horizontal axis. In the solid phases (c, d), the classical density-wave ordered configurations have the maximum weights $\sim \mathcal{O}(10^{-2})$; the magnitudes of these peaks are much larger than any in the disordered or liquid regimes. The presence or absence of ordering is also registered in the associated static structure factors (right).

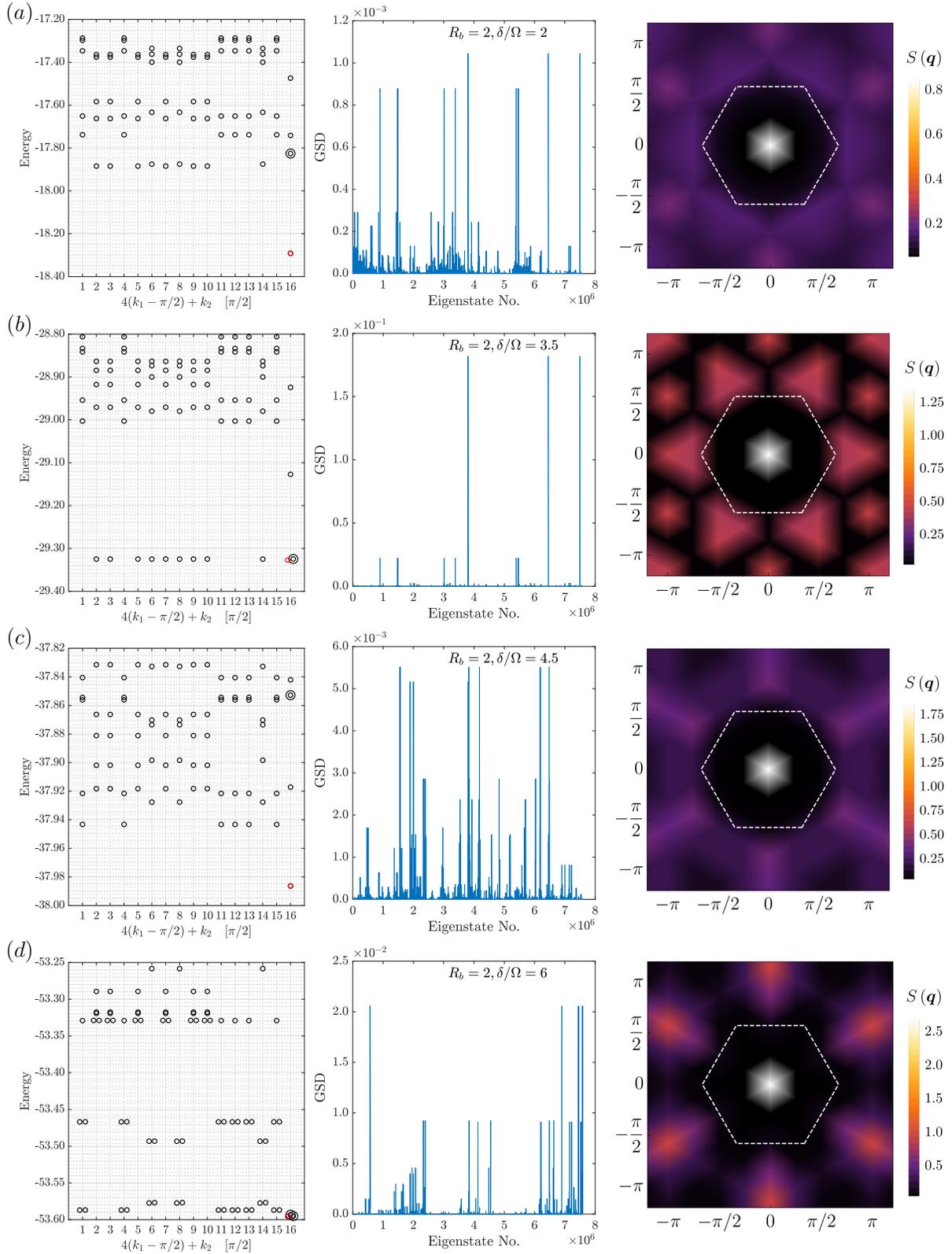

FIG. S8. **Excitation spectra and ground-state properties at** $R_b = 2.0$. Here, we plot the same quantities as in Fig. S7, but for a higher blockade radius. On increasing detuning, the sequence of phases now encountered in going from the disordered to the ordered side is slightly different: from the staggered phase (b), the system enters the liquid regime (c), followed by a transition into the string phase(d). This suggests a scenario where the tip of the staggered lobe bends downwards in the phase diagram and is briefly intersected by the $R_b = 2.0$ line. All the features discussed previously for the liquid state remain unaltered. Note that the structure factor of the staggered phase shown here appears different from that in Fig. 3(c); the former is the sum of $C_3$-rotated copies of the latter.



\end{document}